\documentclass[12pt,prd]{revtex4}

\usepackage{amsmath}
\usepackage{graphicx,color}

\newcommand{\id}{{\bf 1}}
\newcommand{\arxiv}[1]{\href{http://arxiv.org/abs/#1}{arXiv:#1}}

\begin{document}

\hfill\parbox{4cm}
{
 BNL-94327-2010-JA\\
 YITP-10-93\\
}

\title{
Index Theorem and Overlap Formalism with\\
Naive and Minimally Doubled Fermions 
}

\author{Michael Creutz}
\email{creutz@bnl.gov}
\affiliation{Physics Department, Brookhaven National Laboratory, 
Upton, NY 11973, USA}

\author{Taro Kimura}
\email{kimura@dice.c.u-tokyo.ac.jp}
\affiliation{Department of Basic Science, University of Tokyo,
         Tokyo 153-8902, Japan}

\author{Tatsuhiro Misumi}
\email{misumi@yukawa.kyoto-u.ac.jp}
\affiliation{Yukawa Institute for Theoretical Physics, Kyoto University,
         Kyoto 606-8502, Japan}

\begin{abstract}
We present a theoretical foundation for the Index theorem in naive and 
minimally doubled lattice fermions by studying the spectral flow
of a Hermitean version of Dirac operators. We utilize the point 
splitting method to implement flavored mass terms, which play an 
important role in constructing proper Hermitean operators. We show 
the spectral flow correctly detects the index of the would-be zero 
modes which is determined by gauge field topology. Using the flavored
mass terms, we present new types of overlap fermions from the naive 
fermion kernels, with a number of flavors that depends on the choice of
the mass terms. We succeed to obtain a single-flavor naive overlap 
fermion which maintains hypercubic symmetry. 
\end{abstract}

\maketitle

\newpage

%%%%%%%%%%   Introduction   %%%%%%%%%%

\section{Introduction}
\label{sec:Intro}

Studies of topological aspects have been attracting a great deal of
attention in lattice QCD. Issues here include whether the index
theorem relating gauge field topology and fermionic zero modes is well
defined.  Lattice fermion actions without doublers do illustrate the
index theorem nicely.  For example, one old-fashioned way to study
topology in QCD is to use Wilson fermion \cite{Wil, SmtVnk}, although it
requires fine-tuning of a mass parameter because of explicit breaking
of chiral symmetry.  The overlap fermion \cite{GW, Neu}, which is
built on the Wilson fermion kernel, resolves this problem since it
possesses both exact chiral symmetry and exact zero modes related to
gauge field topology, agreeing with the index theorem
\cite{EdHelNar, AdamsInd}. The domain wall formalism \cite{Kap, FuSh}, 
closely related to the overlap fermion, is also a useful tool to study
topological effects in QCD. On the other hand, the index is hidden in
lattice fermions with species doublers since the index effect cancels 
between doubling pairs. Although a theoretical approach to the index 
of the staggered fermions \cite{KS, Suss, Sha} were developed 
in \cite{SmtVnk, SmtVnk2}, it does not give an integer value from 
the beginning and requires a renormalization depending on the full 
ensemble of the gauge fields. Thus it is not easy to define the index 
theorem in naive, staggered and minimally-doubled fermions and thereby 
study topological effects.

Ref.~\cite{Adams1} recently presented how to identify the would-be
zero modes and their chiralities with staggered fermions
away from the continuum limit using the spectral flow of
a certain Hermitean version of the Dirac operator. The integer index
obtained here correctly illustrates the gauge topological charge up to
a factor coming from tastes. In a sequent paper \cite{Adams2}, the
author presents a new version of the overlap fermion, the ``staggered
overlap fermion'' constructed from the staggered kernel. It is a
two-flavor overlap fermion in four dimensions and the associated index
of exact zero modes illustrates the gauge field topology correctly
again. In Ref.~\cite{Hoel} the possibility to construct a
single-flavor version of the staggered overlap fermion is also
discussed.  The key for theoretical foundation of the index theorem
and the sequent overlap formalism here is to introduce a taste
sensitive mass which is essential to reveal the hidden index with the
spectral flow of the Hermitean operator.  The new mass term, which
assigns positive and negative masses to tastes depending on their
flavor-chiralities, removes the cancellation of the index between
tastes.  When the staggered kernel with this kind of mass is
substituted into the overlap formalism, some tastes with positive mass
decouple while other tastes with negative mass yield massless modes,
which correspond to the two-flavor staggered overlap fermion. It is a
universal feature for fermions with species doublers that you can
obtain the associated overlap fermion with a proper flavored mass term
illustrating the correct index.

One natural extension of this approach is to apply it to the naive
lattice fermion.  As is well-known, the naive fermion includes 16
species while staggered has 4 species in four dimension thus it is
expected we can take a parallel process to establish the integer index 
theorem without renormalization and construct overlap fermion with 
the naive kernel. However one
obstacle in this case is the individual species appear most easily
only in momentum space, which leads to difficulty of identifying them
as flavors and introducing flavored mass terms.  We have a similar
problem with minimally doubled fermions 
\cite{Sha, Kar, Wilc, MD1, MD2, VanSmt, SmtGlt}, where there are two species 
in momentum space. So far there has been no theoretical foundation of 
the index theorem for naive and minimally doubled fermions \cite{Cha}.

In this paper, we successfully identify species in naive and minimally
doubled fermions as flavors by using the point splitting method, which
is proposed in Ref.~\cite{CreutzPS} to define up and down quark fields
in minimally doubled fermions. By this identification, we define
proper flavored mass terms to extract the index in the spectral flow
of the associated Hermitean version of the Dirac operator. Here we
follow a parallel approach for identifying the would-be zero modes and
their chiralities to that proposed in the staggered case
\cite{Adams1}.  Then we find the spectral flow correctly illustrates
the index determined by the gauge field topology in naive and
minimally doubled fermions.  In addition we present new versions of
overlap fermions built from the naive fermion kernel with added
flavored mass terms. Especially we construct a single-flavor naive
overlap fermion by choosing a certain flavored mass which assigns
negative mass to one and positive mass to all other species.  One good
property of this fermion over the minimally doubled case is it
maintains hypercubic symmetry.  Furthermore it appears to have more
stability of the index against disorder than the Wilson overlap
fermion. Thus this overlap fermion might be more suitable for
practical simulations.

%%%%%%%%%%  Point splitting and flavored mass terms   %%%%%%%%%%

\section{Point splitting and flavored mass terms}
\label{sec:PS}

With the $d$-dimensional naive lattice fermion there are $2^{d}$ species
and the associated massless Dirac operator is anti-Hermitean.  These
properties are common with the staggered fermion except that there are
only $2^{d/2}$ tastes in the latter case. Thus it seems easy to apply
the approach for the index of staggered fermion \cite{Adams1} to the
naive fermion. In this spectral-flow approach the flavored mass term 
is essential, which gives $+$ and $-$ masses to flavors(species) 
depending on their chiral charges. However the species in the naive fermion
appear naturally only in momentum space. Therefore in this case we
cannot implement flavored mass terms straightforwardly since we need
to identify species in momentum space as independent
flavors. To resolve this problem we apply a method called ``point
splitting,'' originally proposed to define up and down quark
fields in minimally doubled fermion \cite{CreutzPS}. By using this
method we obtain flavored mass terms to give proper Hermitean
operators for the spectral flow both in minimally doubled and naive
fermions. In this section we first introduce the point splitting for
minimally doubled fermions and implement flavored mass terms. Secondly
we perform the parallel approach for naive fermions.

\subsection{Minimally doubled action}
\label{sec:MDA}

Here we introduce the point splitting method to obtain flavored mass terms in 
minimally doubled fermions. We focus only on the Karsten-Wilczek type 
\cite{Kar, Wilc} all through this paper. Now we begin with the introduction 
of minimally doubled fermions. The $d=4$ Karsten-Wilczek action is obtained 
by introducing a Wilson-like term proportional to $i\gamma_4$. 
Its position-space expression is
\begin{eqnarray}
 S_{\mathrm{md}} & = &
  \sum_{x} 
  \Bigg[
   \frac{1}{2} \sum_{\mu=1}^3 \bar\psi_x \gamma_\mu
   \left(
    U_{x,x+e_\mu} \psi_{x+e_\mu} - U_{x,x-e_\mu}\psi_{x-e_\mu}
   \right)
   \nonumber \\
 & & + \frac{i}{\sin \alpha}
  \left(
   (\cos \alpha + 3) \bar\psi_x \gamma_4 \psi_x
   - \frac{1}{2} \sum_{\mu=1}^4 \bar \psi_x \gamma_4
   \left(
    U_{x,x+e_\mu} \psi_{x+e_\mu} + U_{x,x-e_\mu} \psi_{x-e_\mu}
   \right)
  \right)
  \Bigg]
\end{eqnarray}
where the link variables satisfy $U_{xy}=U_{yx}^\dag$. For the free theory, 
the associated Dirac operator in momentum space is given by
\begin{equation}
 D_{\mathrm{md}}(p) =
  i \sum_{\mu=1}^3 \gamma_\mu \sin p_\mu
  + \frac{i\gamma_4}{\sin \alpha}
  \left(
   \cos \alpha + 3 - \sum_{\mu=1}^4 \cos p_\mu
  \right),
\end{equation}
where the parameter $\alpha$ adjusts the relative positions of zeros.
It has only two zeros located at $p=(0, 0, 0, \pm \alpha)$.  These two
species are not equivalent since the gamma matrices are differently
defined between them as $\gamma_\mu'=\Gamma^{-1} \gamma_\mu \Gamma$.
In the above case the transformation matrix is given by 
$\Gamma = i \gamma_4 \gamma_5$.  This means the chiral symmetry 
possessed by this action is identified as a
flavored one given by $\gamma_5\otimes\tau_3$.

The point splitting identifies these inequivalent species as
independent flavors.  In this method each flavor field is defined so
that the associated propagator includes only a single pole. Thus we
point-split the original fermion field by multiplying factors removing
another undesired pole in momentum space,
\begin{equation}
 u(p-\alpha e_4) = \frac{1}{2} 
  \left(
   1 + \frac{\sin p_4}{\sin \alpha}
  \right)
  \psi(p),
\end{equation}
\begin{equation}
 d(p+\alpha e_4) = \frac{1}{2} \Gamma
  \left(
   1 - \frac{\sin p_4}{\sin \alpha}
  \right)
  \psi(p).
\end{equation}
Here $u(p-\alpha e_4)$ and $d(p+\alpha e_4)$ fields correspond to the
poles at $p=(0, 0, 0, \alpha)$ and $(0, 0, 0, -\alpha)$, respectively.
We remark both of them yield single fermionic modes and the associated
point-split fields in position space are composed of the original field 
and the two kinds of nearest neighbors \cite{CreutzPS}. With these flavor fields 
we obtain a flavor-multiplet field as following,
\begin{equation}
 \Psi(p) = 
  \left(
   \begin{array}{c}
    u(p-\alpha e_4) \\ d(p+\alpha e_4)
   \end{array}
  \right).
\end{equation}
Here $\gamma_5$ multiplication on the original Dirac field is identified as
\begin{equation}
 \gamma_5 \psi(p)
  \quad \longrightarrow \quad
  \left(
    \begin{array}{cc}
     + \gamma_5 & \\
     & - \gamma_5 \\
    \end{array}
  \right) \Psi(p)
   = \left(\gamma_5 \otimes \tau_3\right) \Psi(p).
   \label{gamma5_MD}
\end{equation}
Here we introduce a multiplet representation as a direct product 
of the Pauli matrix to express the 2-flavor structure.
It means the flavored chiral symmetry generated by $\gamma_5\otimes\tau_3$
is exactly preserved in terms of the flavor multiplet while the
flavor singlet chiral symmetry given by $\gamma_5\otimes\id$ is broken 
in higher order terms. It is also the case of the staggered fermion \cite{SmtGlt}.

Now we can introduce a flavor breaking mass into minimally doubled
fermions by a parallel approach to the case of staggered fermions in
Refs.~\cite{Adams1,Adams2}.  We have to choose a flavored mass term so
that it reveals the hidden index when looking into the spectral flow
of a certain Hermitean version of the Dirac operator
$H(m)=\gamma_5(D-M)$, which we will explain in detail in
Sec.~\ref{sec:SP}. (This $\gamma_5$ is regarded as
$\gamma_5\otimes\tau_3$ in terms of the flavor multiplet as shown in
(\ref{gamma5_MD}).)  We need to impose that $\gamma_{5}M$ should be
flavor-singlet as $\gamma_5 M = \gamma_5\otimes\id$. Thus the
corresponding flavored mass term here is
$M=m\left(\id\otimes\tau_3\right)$ with a mass parameter $m$.  To
implement this flavored mass term into the action, we rewrite the mass
term with the original Dirac field as
\begin{equation}
 \bar\Psi(p) \left(\id\otimes\tau_3\right) \Psi(p)
  = \bar uu(p-\alpha e_4) - \bar dd(p+\alpha e_4)
  = \frac{\sin p_4}{\sin \alpha} \bar \psi(p)\psi(p).
  \label{mass01_MD}
\end{equation}
It is straightforward to obtain the mass term in the position space
with the link variables present,
\begin{equation}
 M_{\mathrm{md}} = \frac{m}{2i\sin \alpha} \bar \psi_x
  \left(
   U_{x,x+e_4} \psi_{x+e_4} - U_{x,x-e_4} \psi_{x-e_4}
  \right).
\label{2dMMass}
\end{equation}
Here we also write a flavor-singlet mass term
\begin{equation}
 \bar\Psi(p) \left(\id\otimes\id\right) \Psi(p)
  = \bar uu(p-\alpha e_4) + \bar dd(p+\alpha e_4)
  = \frac{1}{2} \left(1+\frac{\sin^2 p_4}{\sin^2 \alpha}\right) \bar
  \psi(p)\psi(p),
  \label{mass02_MD}
\end{equation}
which is not required for the study of the index.  The associated massive
Dirac operator $D_{\mathrm{md}} - M_{\mathrm{md}}$ is non-Hermitean,
and when gauge fields are present the mass term does not commute with
the kinetic term $[D_{\mathrm{md}}, M_{\mathrm{md}}]\not=0$. Thus 
the Dirac operator
eigenvalues are complex. Indeed it is essential for the
purpose to detect the index from the spectral flow of the Hermitean
operator since it relies on real eigenvalues of the Dirac operator.

\begin{figure}
 \begin{center}
  \includegraphics[width=15em]{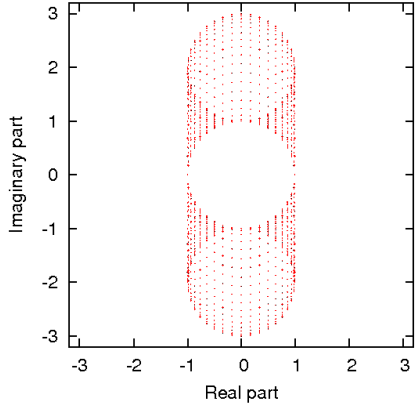}
 \end{center}
 \caption{Complex spectrum of the free non-Hermitean Dirac operator
   $D_{\mathrm{md}}-M_{\mathrm{md}}$ for the $d=2$ free field case on
   a $36 \times 36$ lattice with a mass parameter $m=1$. The spectrum
   is split into two branches crossing the real axis at $|m|$.}
 \label{Dirac_ev_KW}
\end{figure}

In Fig.~\ref{Dirac_ev_KW} we show a numerical result of complex
eigenvalues of the Dirac operator for the $d=2$ free case with a
parameter $\alpha=\pi/2$. Here the low-lying spectrum is split into 
two branches crossing the real axis at the magnitude of the mass parameter $|m|$.
It means that the flavored mass $-M_{\mathrm{md}}=\mathrm{diag}(-m,+m)$ assigns 
$-m$($+m$) to modes depending on $+1$($-1$) chiral charges, or equivalently 
$+1$($-1$) eigenvalues for $\id\otimes\tau_{3}$. In other words the flavored 
mass term splits the minimally doubled fermion into two single Dirac fermions 
with $-m$ and $+m$ as in the staggered case. In Sec.~\ref{sec:SP} we will see 
this flavored mass gives the spectral flow of the Hermitean operator illustrating 
the correct index related to the gauge topology.

\subsection{Naive action}
\label{sec:NA}

We can apply the same approach to the naive lattice fermion \cite{Wil} to
obtain a proper flavored mass term.
The action of the naive fermions in general dimensions is simply given by
\begin{equation}
 S_{\mathrm{n}} = \frac{1}{2} \sum_x \sum_{\mu=1}^d \bar\psi_x \gamma_\mu
  \left(
   U_{x,x+e_\mu} \psi_{x+e_\mu} - U_{x,x-e_\mu} \psi_{x-e_\mu}
  \right).
\end{equation}
For simplicity here we consider the $d=2$ naive fermions.  The Dirac
operator has four zeros $(0, 0)$, $(\pi, 0)$, $(0, \pi)$ and
$(\pi,\pi)$ in momentum space, thus we introduce four associated
point-split fields
\begin{eqnarray}
 \psi_{(1)}(p-p_{(1)}) & = & \frac{1}{4} (1+\cos p_1)(1+\cos p_2)
  \Gamma_{(1)} \psi(p),
  \nonumber \\
 \psi_{(2)}(p-p_{(2)}) & = & \frac{1}{4} (1-\cos p_1)(1+\cos p_2)
  \Gamma_{(2)} \psi(p),
  \nonumber \\
 \psi_{(3)}(p-p_{(3)}) & = & \frac{1}{4} (1+\cos p_1)(1-\cos p_2)
  \Gamma_{(3)} \psi(p),
  \nonumber \\
 \psi_{(4)}(p-p_{(4)}) & = & \frac{1}{4} (1-\cos p_1)(1-\cos p_2)
  \Gamma_{(4)} \psi(p),
\end{eqnarray}
whose locations of zeros, chiral charges and transformation matrices 
$\Gamma$ giving the corresponding set of gamma matrices,
$\gamma_\mu^{(i)}=\Gamma_{(i)}^\dag \gamma_\mu \Gamma_{(i)}$, are listed in 
Table \ref{gammas_naive2d}. The flavor-multiplet field is given by
\begin{equation}
 \Psi(p) =
  \left(
   \begin{array}{c}
    \psi_{(1)}(p-p_{(1)}) \\ \psi_{(2)}(p-p_{(2)}) \\
    \psi_{(3)}(p-p_{(3)}) \\ \psi_{(4)}(p-p_{(4)})
   \end{array}
   \right).
\end{equation}   
\begin{table}[tbp]
 \begin{center}
{\tabcolsep = 1em
  \begin{tabular}{cccc} \hline\hline
   label & position & $\chi$ charge & $\Gamma$ \\ \hline
   $1$ & $(0, 0)$ & $+$ & $\id$ \\
   $2$ & $(\pi, 0)$ & $-$ & $i\gamma_1\gamma_5$ \\
   $3$ & $(0, \pi)$ & $-$ & $i\gamma_2\gamma_5$ \\
   $4$ & $(\pi, \pi)$ & $+$ & $\gamma_5$ \\ \hline\hline
  \end{tabular}
}
 \end{center}
 \caption{Chiral charges and transformation matrices for each of zeros in the $d=2$
 naive fermions with $\gamma_1 = \sigma_1$, $\gamma_2=\sigma_2$ and $\gamma_5=\sigma_3$.}
 \label{gammas_naive2d}
\end{table}
Here the operation of $\gamma_5$ on the original fermion 
field again means the flavored chiral transformation in the sense of the flavor multiplet as
\begin{equation}
 \gamma_5 \psi(p)
  \quad \longrightarrow \quad
  \left(
   \begin{array}{cccc}
    +\gamma_5 & & & \\
    & -\gamma_5 & & \\
    & & -\gamma_5 & \\
    & & & +\gamma_5 \\
   \end{array}
  \right) \Psi(p)
  = \left(\gamma_5\otimes\left(\tau_3\otimes\tau_3\right)\right) \Psi(p).
\end{equation}
Here we introduce a multiplet representation as two direct products 
of the Pauli matrix to express the 4-flavor structure.
Now we adopt the criterion for the proper flavored mass term that 
$\gamma_{5}M$ should be flavor-singlet as $\gamma_5 M = \gamma_5\otimes\id$.
Therefore the desirable flavored mass for the naive fermion is given by
\begin{equation}
 \bar\Psi(p) \left(\id\otimes\left(\tau_3\otimes\tau_3\right)\right) \Psi(p)
  = \cos p_1 \cos p_2 \bar\psi(p) \psi(p).
\label{MN}
\end{equation}
Please note there are also other kinds of flavored mass terms in this case:
\begin{equation}
 \bar\Psi(p) \left(\id\otimes\left(\tau_3\otimes\id\right)\right) \Psi(p)
  = \frac{1}{2} \cos p_1 (1+\cos^2 p_2)  \bar\psi(p) \psi(p),
\end{equation}
\begin{equation}
 \bar\Psi(p) \left(\id\otimes\left(\id\otimes\tau_3\right)\right) \Psi(p)
  = \frac{1}{2} (1+\cos^2 p_1) \cos p_2  \bar\psi(p) \psi(p),
\end{equation}
\begin{equation}
 \bar\Psi(p) \left(\id\otimes\left(\id\otimes\id\right)\right) \Psi(p)
  = \frac{1}{4} (1+\cos^2 p_1) (1+\cos^2 p_2)  \bar\psi(p) \psi(p).
\end{equation}
These varieties are not useful to detect the index via the spectral
flow of the Hermitean operator. However we will show they are
essential when we search for a single-flavor naive overlap fermion in
Sec.~\ref{sec:OF}.  All of these mass terms are spread over several
nearby sites, and therefore their position-space expressions include
hopping terms with associated gauge field matrices. It is convenient
to introduce the usual translation operators
$T_{\pm\mu}\psi_x=U_{x,x\pm e_\mu}\psi_{x\pm e_\mu}$ and $C_\mu =
(T_{+\mu} + T_{-\mu})/2$.  Then the flavored mass for the index
(\ref{MN}) is written in the position space as
\begin{equation}
 M_{\tau_3\otimes\tau_3} = m \sum_{sym.} C_1 C_2
 \equiv M_{\mathrm{n}},
\label{2dNMass}
\end{equation}
where $\sum_{sym.}$ stands for symmetric summation over the order of
the factors $C_1$ and $C_2$, and $m$ stands for a mass
parameter. We can also write the extra mass terms in the position space as
\begin{equation}
 M_{\tau_3\otimes\id} = \frac{m_{\tau_3\otimes\id}}{2} \sum_{sym.}
  \left(1+C_1^2\right) C_2,
\label{exM1}
\end{equation}
\begin{equation}
 M_{\id\otimes\tau_3} = \frac{m_{\id\otimes\tau_3}}{2} \sum_{sym.}
  C_1 \left(1+C_2^2\right),
\label{exM2}
\end{equation}
\begin{equation}
 M_{\id\otimes\id} = \frac{m_{\id\otimes\id}}{4} \sum_{sym.}
  \left(1+C_1^2\right)\left(1+C_2^2\right).
\label{exM3}
\end{equation}
Here we comment on the possibility that the two-step hoppings 
in the same directions would be reduced to zero-step terms with
reduction in the effects of gauge field fluctuations.

\begin{figure}
 \begin{center}
  \includegraphics[width=15em]{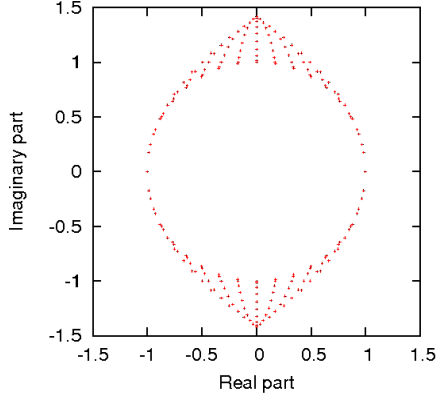}
 \end{center}
 \caption{Complex spectra of non-Hermitean Dirac operator
 $D_{\mathrm{n}}-M_{\tau_3\otimes\tau_3}$ for the $d=2$ free field case
 on a $36 \times 36$ lattice with mass parameter
 $m=1$.The spectrum is split into two doubled branches crossing 
 the real axis at $|m|$.}
 \label{Dirac_ev_n2d}
\end{figure}

Now let us look into the eigenvalues of the $d=2$ naive Dirac operator
$D_{\mathrm{n}}-M_{\mathrm{n}}$ with the flavored mass
$M_{\mathrm{n}}$ in Eq.~(\ref{2dNMass}).  In Fig.~\ref{Dirac_ev_n2d}
we show a numerical result of the complex eigenvalues. Here the low-lying
spectrum is again split into two branches crossing the real axis at
the magnitude of the mass parameter $|m|$.  However in this case both
of the two branches are doubled.  (We will be convinced of this
doubling in Fig.~\ref{Dirac_ev_n2d02} where it is lifted by the other
mass terms.) It means that the flavored mass term
$-M_{\mathrm{n}}=\mathrm{diag}(-m,+m,+m,-m)$ assigns $-m$($+m$) to
modes depending on $+1$($-1$) chiral charges, or $+1$($-1$)
eigenvalues for $\id\otimes(\tau_{3}\otimes\tau_{3})$. Thus the $d=2$
naive fermion with 4 species is split into two pairs of Dirac fermions
with $-m$ and $+m$. In the next section we will show the spectral flow
of the associated Hermitean operator gives the correct index related
to the gauge topology.

In the $d=4$ case we can apply the same approach to obtain a proper flavored mass term.
In this case there are more possibilities for flavored mass.
So we will not discuss details here, and just show the desirable
flavored mass term to reveal the index.
It is given in the momentum space by  
\begin{equation}
\bar\Psi
   \left(
    \id\otimes\left(\tau_3\otimes\tau_3\otimes\tau_3\otimes\tau_3\right)
   \right)\Psi
   = 
   \left(
    \prod_{\mu=1}^4 \cos p_\mu
   \right) \bar\psi\psi ,
\label{4dM1}
\end{equation}
and in the position space by
\begin{equation}
 M_{\mathrm{P}} = m_{\mathrm{P}} \sum_{sym.} 
  \prod_{\mu=1}^4 C_\mu.
\label{4dM2}
\end{equation}
Appendix \ref{sec:N4d} is devoted to details of the flavored mass terms in
$d=4$ naive fermions.  In Fig.~\ref{Dirac_ev_n4d}(a) for the eigenvalues
of the Dirac operator we can verify the modes are split into two
8-degenerate branches, which means the flavored mass decomposes the
$4$d naive fermions into two pairs with negative and positive masses.

%%%%%%%%%%  Spectral flow and Index theorem   %%%%%%%%%%

\section{Spectral flow and the index theorem}
\label{sec:SP}

In this section we obtain the integer index related with gauge field
topology for the naive and minimally doubled fermions.  As in the
cases of Wilson \cite{EdHelNar} and staggered fermions \cite{Adams1,
  Adams2} we utilize the spectral flow of the Hermitean operators with
the flavored mass terms introduced in Sec.~\ref{sec:PS}. Here the
would-be zero modes of the Dirac operators are identified as low-lying
crossings of the eigenvalue flow of the Hermitean operators.

Let us begin with explaining what the spectral flow is.  In the
continuum field theory the index is defined as the difference between
the numbers of zero modes of the massless Dirac operator with positive
and negative chirality, $n_+$ and $n_-$.  The statement of the index
theorem is that the index is just equal to a topological charge $Q$ of
a background gauge configuration up to a sign factor depending on its
dimensionality,
\begin{equation}
 n_+ - n_- = (-1)^{d/2} Q.
\end{equation}
Here the question is how to obtain the index of the Dirac operator.
We can of course calculate the zero-mode chiralities
straightforwardly, but there is a useful way called {\it spectral
  flow}.  To introduce it we first introduce a certain Hermitean
version of the Dirac operator
\begin{equation}
 H(m) = \gamma_5 (D - m).
\end{equation}
where any zero modes of the Dirac operator with $\pm$ chirality
correspond to some eigenmodes of this Hermitean operator with
eigenvalues $\lambda(m)=\mp m$.  If we now consider the flow of the
eigenvalues $\lambda(m)$ as the mass varies, those corresponding to
zero modes will cross the origin with slopes $\mp 1$ depending on
their $\pm$ chirality.  The non-zero eigenmodes of $D$, in contrast,
occur in pairs which are mixed by $H$ and cannot cross zero.
Therefore the index of the Dirac operator is given by minus the
spectral flow of the Hermitean operator, which stands for the net
number of eigenvalues crossing the origin, counted with sign $\pm$
depending on the slope.

The index with lattice Wilson fermions \cite{EdHelNar} can similarly be
obtained from the spectral flow, which in this case means the net
number of eigenvalues crossing zero at low-lying values of $m$,
counted with signs of the slopes.  In the continuum limit, we are only
interested in the crossings at small mass; the massive doublers also
eventually cross zero, but only for large values of $m$.
However for lattice fermions with massless species doublers the index 
cancels between pairs, so that an eigenvalue flow with a simple mass 
term does not properly capture gauge field topology.  This difficulty 
has been recently resolved by introducing a proper flavored (``tasted'') 
mass term in staggered fermions \cite{Adams1,Adams2}.  
The associated Hermitean operator is given by
\begin{equation}
 H_{\mathrm{st}}(m)
 = \Gamma_{55} \left( D_{\mathrm{st}} - m \Gamma_{55} \Gamma_5 \right)
 = \Gamma_{55} D_{\mathrm{st}} - m \Gamma_5,
 \label{Hermitean_st01}
\end{equation}
where $\Gamma_{55}$ and $\Gamma_{5}$ stand for
$\Gamma_{55}=\gamma_5\otimes\gamma_5$ and $\gamma_5\otimes\id$ with
discretization error $\mathcal{O}(a)$ in terms of a taste multiplet.
Indeed it was shown that the index of the staggered fermion is given by
minus the spectral flow of this Hermitean operator. And it was also
pointed out that it illustrates correctly the gauge topology up to a
integer factor from the number of tastes as
$\mathrm{Index}(D_{\mathrm{st}}) = 2^{d/2} (-1)^{d/2} Q$.  Here the
spectral flow again means the net number of eigenvalues crossing zero
at low-lying values of $m$, counted with signs of the slopes.  Thus
the theoretical foundation of the index theorem with staggered
fermions is established without a renormalization depending on the 
gauge ensemble. Here the mass part in the Hermitean operator
Eq.~(\ref{Hermitean_st01}) is approximately proportional to flavor-singlet
gamma-5. Note we constructed flavored mass terms for our cases in the
previous section following this.

Now we can symbolically write a formula for the index as
\begin{equation}
{\rm Index}(D) = -{\rm Spectral\,\,\,flow}(H).
\label{Formula}
\end{equation}
It is quite natural to consider whether this formula is also available to
detect the index of minimally doubled and naive fermions with the
flavored mass terms we proposed.  We will from now show this spectral
flow method can be also applied to these cases.
The associated Hermitean operators for minimally doubled and
naive fermions are given by
\begin{equation}
 H_{\mathrm{md}}(m) = \gamma_5 (D_{\mathrm{md}} - M_{\mathrm{md}}),
  \label{Hermitean_md01}
\end{equation}
\begin{equation}
 H_{\mathrm{n}}(m) = \gamma_5 (D_{\mathrm{n}} - M_{\mathrm{n}}),
  \label{Hermitean_naive01}
\end{equation}
where the matrix $\gamma_5$ is regarded as a flavored one,
$\gamma_5\otimes\tau_3$ for minimally doubled fermions and
$\gamma_5\otimes\left(\tau_3\otimes\tau_3\right)$ for two dimensional naive fermions 
in terms of the flavor multiplet as shown in Sec.~\ref{sec:PS}.
The desirable flavored mass terms $M_{\rm md}$ and $M_{\rm n}$ for $d=2$ have been
already given in Eq.~(\ref{2dMMass}) and (\ref{2dNMass}).
For now we focus on the two dimensional case.

We then numerically calculate the eigenvalue flows of two dimensional
minimally doubled and naive fermions. We consider background
configurations proposed in \cite{SmtVnk} for the staggered case
\cite{Adams1}: we start with a smooth $U(1)$ gauge field with
topological charge $Q$,
\begin{equation}
 U_{x,x+e_1} = e^{i\omega x_2}, \qquad
 U_{x,x+e_2} = 
 \left\{
  \begin{array}{cl}
   1 & \quad \left( x_2 = 1, 2, \cdots, L - 1 \right) \\
   e^{i\omega L x_1} & \quad \left(x_2 = L\right)
  \end{array}
 \right. ,
\end{equation}
where $L$ is the lattice size and $\omega$ is the curvature given by
$\omega = 2\pi Q$.
Then, to emulate a typical gauge configuration of a practical
simulation, we introduce disorder effects to link variables by random
phase factors, $U_{x,y} \to e^{ir_{x,y}} U_{x,y}$, where $r_{x,y}$ is a
random number uniformly distributed in $[-\delta \pi, \delta \pi]$. The
parameter $\delta$ determines the magnitude of disorder.

%NOTE: We should explicitly define the configuration used here.  Also
%mention that it is a U(1) gauge field (is this right?).  Also comment
%on how the crossings depend on delta; do extraneous ones start to
%appear as delta approaches unity?  

\begin{figure}
 \begin{center}
  \includegraphics[width=20em]{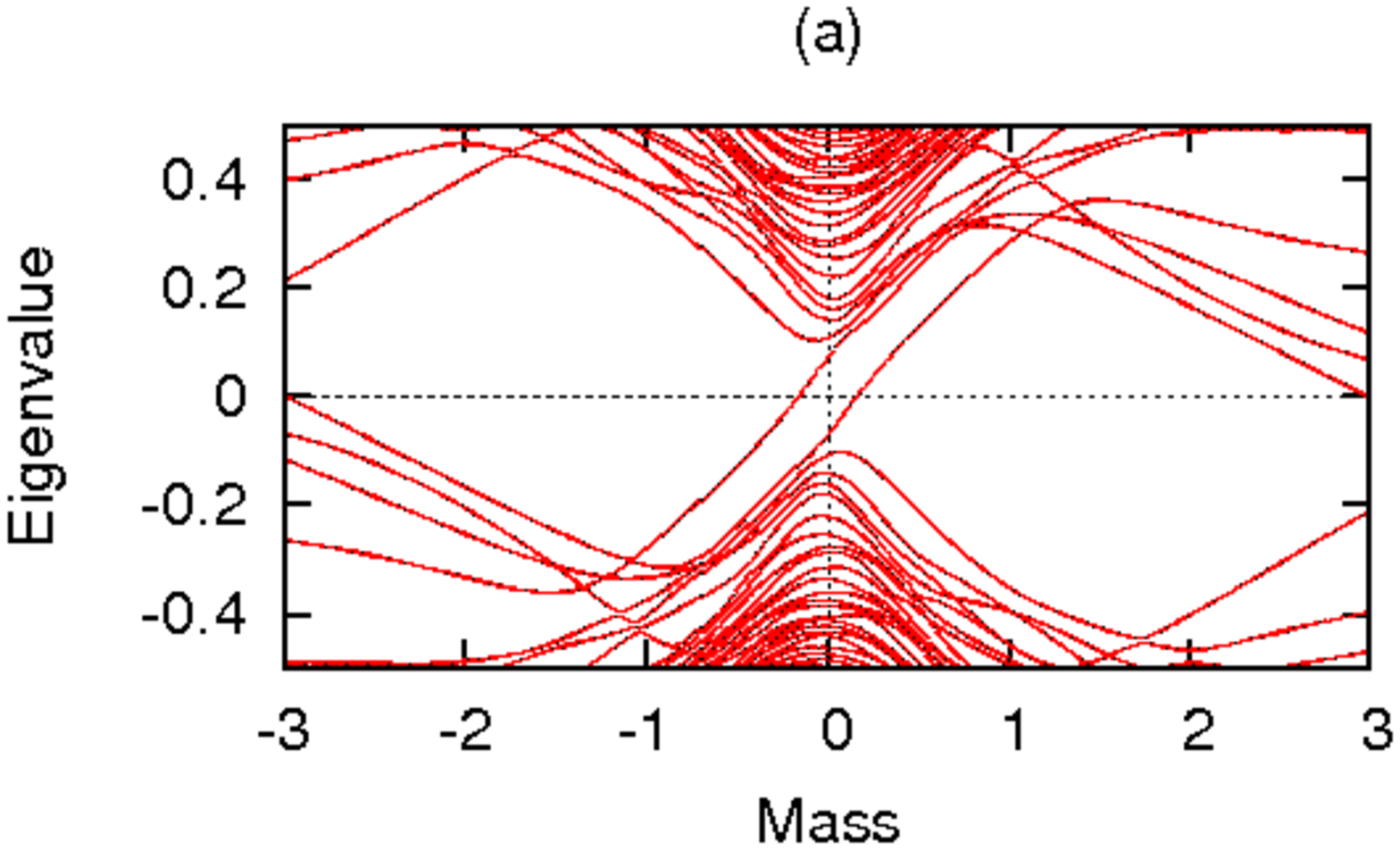} \qquad
  \includegraphics[width=20em]{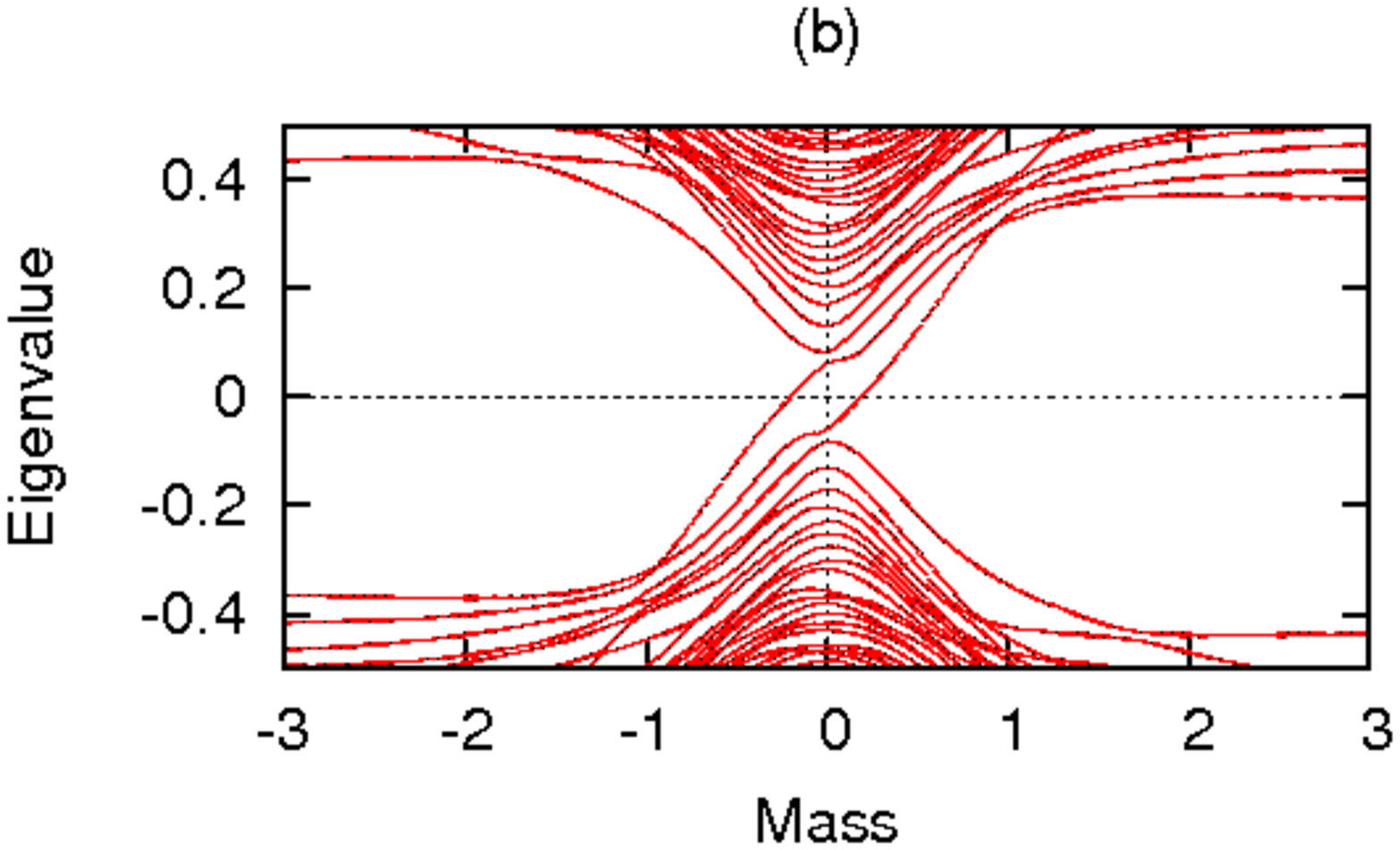}
 \end{center}
 \caption{Spectral flows of (a) Minimally doubled and (b) naive Hermitean
 operators with a $Q=1$, $\delta=0.25$ background configuration on a $16\times 16$
 lattice. Two single crossings with positive slopes are seen in (a), which 
 means the index is $-2$. Two doubled crossings with positive slopes are seen 
 in (b), which means the index is $-4$.}
 \label{sf_pic}
\end{figure}

\begin{figure}
 \begin{center}
  \includegraphics[width=20em]{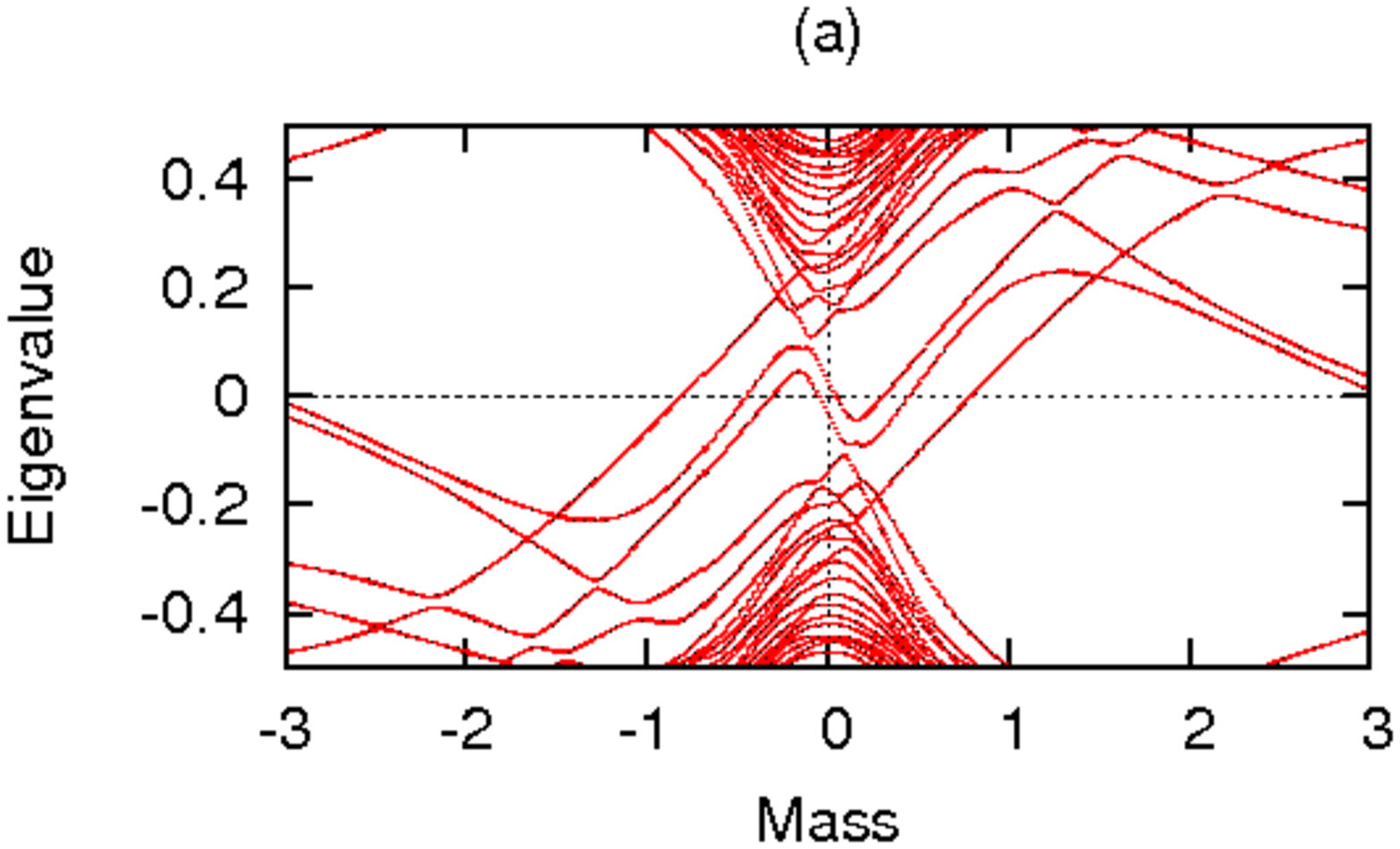} \qquad
  \includegraphics[width=20em]{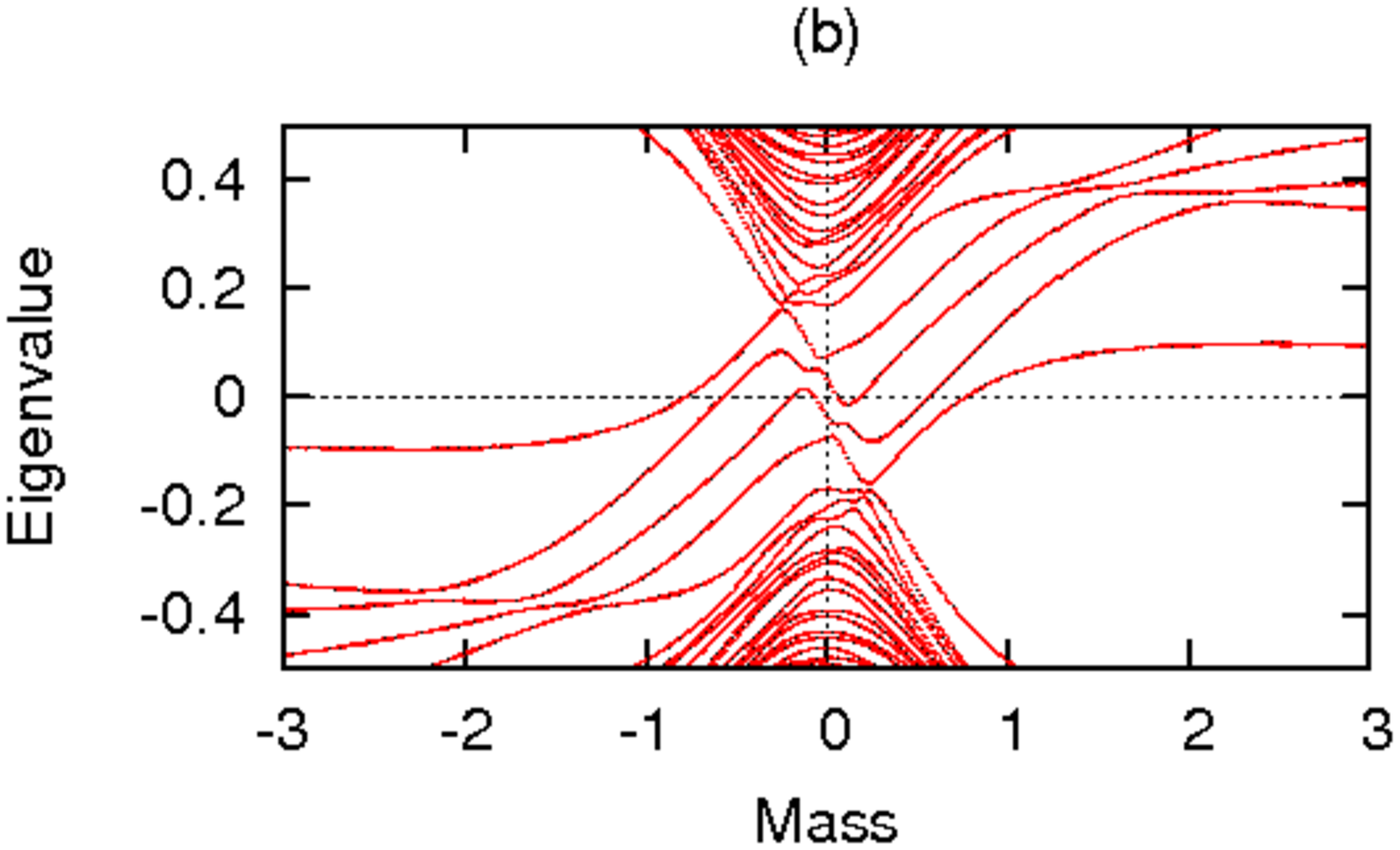}
 \end{center}
 \caption{Spectral flows of (a) Minimally doubled and (b) naive Hermitean
 operators with a $Q=2$, $\delta=0.2$ background configuration on a $16\times 16$
 lattice. Six single crossings with positive slopes and two single crossings 
 with negative slopes are seen in (a), which 
 means the index is $-4$. Six doubled crossings with positive slopes and two 
 doubled crossings with negative slopes are seen 
 in (b), which means the index is $-8$.}
 \label{sf_pic2}
\end{figure}

Fig.~\ref{sf_pic}(a) shows the eigenvalue flow of the minimally
doubled Hermitean operator (\ref{Hermitean_md01}).  It is calculated
with a $Q=1$ and $\delta=0.25$ background configuration on a $16
\times 16$ lattice.  There are two low-lying crossings around $m=0$
with positive slopes, which correspond to would-be zero modes.  With
the formula (\ref{Formula}), it means the index of the Dirac operator
of the minimally doubled fermion in this case is $-2$.  This result is
consistent with the index theorem for the minimally doubled fermions
given by
\begin{equation}
 \mathrm{Index}(D_{\mathrm{md}}) = 2 (-1)^{d/2} Q,
\label{MDtheorem}
\end{equation}
which contains a factor 2 reflecting two species.  This relation is
also satisfied by cases with other topological charges, as shown in
Fig.~\ref{sf_pic2}(a) for the case for $Q=2$.  Here the net number of
crossings counted with $\pm$ depending on the slopes is $4$.  It means
the corresponding index is $-4$, which is consistent with
(\ref{MDtheorem}).  We also emphasize that there is a clear separation
between low- and high-lying crossings in Fig.~\ref{sf_pic}(a) where
low-lying ones are localized about $m=0$ and high-lying ones are
located at large $|m|$. Although further numerical study with realistic 
gauge configurations is required, it indicates the zero modes 
and the index tend to be robust against randomness of the gauge configuration. 
Now we have established the index theorem with minimally doubled fermions.

Next results for the naive fermion case are shown in
Fig.~\ref{sf_pic}(b).  The calculation is done with the same
background configuration as the minimally doubled case.
Fig.~\ref{sf_pic}(b) shows the eigenvalue flow of the naive Hermitean
operator (\ref{Hermitean_naive01}).  There are two doubled crossings
around the origin: Here we can verify they are doubled and there
totally exist four crossings as shown in Fig.~\ref{sf_pic02}(a) by
introducing other kinds of mass terms given in Sec.~\ref{sec:PS}.
%like $m_{\tau_3\otimes\id}/m_{\tau_3\otimes\tau_3} =
%m_{\id\otimes\tau_3}/m_{\tau_3\otimes\tau_3} \ll 1$.  
Again with the formula (\ref{Formula}) we obtain the index of the Dirac
operator of the naive fermion in this case is $-4$. This result satisfies the
index theorem for the naive fermion given by
\begin{equation}
 \mathrm{Index}(D_{\mathrm{n}}) = 2^{d} (-1)^{d/2} Q,
  \label{naive_index_thrm}
\end{equation}
where it contains a factor $2^{d}$ reflecting $2^{d}$ species.
This theorem is also satisfied by the cases with other topological charges,
as shown in Fig.~\ref{sf_pic2}(b) for $Q=2$.
Here the spectral flow is $8$ since all the crossings are doubled in this case.
It means the corresponding index is $-8$, which is consistent with (\ref{naive_index_thrm}).    
There is also large separation between low-lying and high-lying crossings.
It indicates the zero modes and the index tend to be robust as in the case of minimally doubled fermions.
Now we have established the index theorem with the naive fermion.

We remark both of the Hermitean operators, (\ref{Hermitean_md01}) and
(\ref{Hermitean_naive01}), satisfy a relation of $\gamma_5 H(m) \gamma_5
= - H(-m)$, reflecting the fact that the original actions have an exact
chiral symmetry while the Wilson fermion does not. 
We can see low-lying and high-lying crossings get closer as the
randomness effect becomes larger in the both cases, but this original
chiral symmetry actually improves the lower bound of $H^{2}(m)$
satisfying $H^{2}(m)=D^{\dag}D+m^{2}$, which enhances the stability of
the index against disorder. 
Thus overlap formalisms with originally chirally symmetric
fermions would be more applicable to practical simulations as we will discuss later.
In the end of this section, let us comment on $d=4$ cases.
We can perform the same argument for the $4$-dimensional minimally doubled 
fermions and naive fermions. Especially in the case of the $4$d naive fermion,
we can show the integer index is obtained from the spectral flow and it correctly 
illustrates the topological charge in accordance with Eq.~(\ref{naive_index_thrm})
by introducing the flavored mass term
\begin{equation}
 M_{\mathrm{P}} = m_{\mathrm{P}} \sum_{sym.} 
  \prod_{\mu=1}^4 C_\mu,
\label{4dM}  
\end{equation}
In Appendix \ref{sec:N4d} we will explain this mass term in detail and 
show the eigenvalues of the associated Dirac operator in Fig.~\ref{Dirac_ev_n4d}(a).

%%%%%%%%%%   Overlap formalism   %%%%%%%%%%

\section{Overlap formalism}
\label{sec:OF}

In this section we discuss new versions of overlap fermions
constructed from the naive and minimally doubled Dirac kernels with
the flavored mass terms.  The main result here is we obtain a
single-flavor overlap fermion with hypercubic symmetry from the naive
fermion kernel, which may be somewhat simpler than the Wilson overlap fermion.

Firstly we show the index of exact zero modes of the naive and 
minimally doubled overlap fermions also illustrate the topological 
charge correctly. We now introduce minimally doubled and naive 
versions of overlap Dirac operators,
\begin{equation}
 D_{\mathrm{mo}} = 1 + \gamma_5 
  \frac{H_{\mathrm{md}}}{\sqrt{H^2_{\mathrm{md}}}}, \qquad
 D_{\mathrm{no}} = 1 + \gamma_5 
  \frac{H_{\mathrm{n}}}{\sqrt{H^2_{\mathrm{n}}}},
\end{equation}
where these Hermitean operators have been introduced in
(\ref{Hermitean_md01}) and (\ref{Hermitean_naive01}).  These depend on
the mass parameter, which should be chosen in an appropriate region.
Then we can obtain the corresponding Ginsparg-Wilson relations
\begin{equation}
 \left\{\gamma_5, D_{\mathrm{mo}}\right\}
 = D_{\mathrm{mo}} \gamma_5 D_{\mathrm{mo}}, \qquad
 \left\{\gamma_5, D_{\mathrm{no}}\right\}
 = D_{\mathrm{no}} \gamma_5 D_{\mathrm{no}},
 \label{GW_md_naive}
\end{equation}
where, as we have discussed, the $\gamma_{5}$ here can be identified
as a flavored one in terms of the flavor multiplet, for example
$\gamma_{5}\,\,\to\,\,\gamma_{5}\otimes\tau_{3}$ for the minimally
doubled fermion and
$\gamma_{5}\,\,\to\,\,\gamma_{5}\otimes(\tau_{3}\otimes\tau_{3})$ for
the $d=2$ naive fermion. 
%In this sense this Ginsparg-Wilson relation
%is a flavored symmetry. 
%NOTE: with one physical fermion there is no flavor left, so to call 
%this a flavored symmetry is confusing
Note that only half of the original
flavors(species) with negative mass are converted into physical
massless modes in these overlap fermions while the others with
positive mass become massive and decouple in the continuum limit.
This is because
the flavored mass terms we introduced assign negative mass to half of
species and positive mass to the others.  Thus there is only a single
physical mode for the minimally doubled overlap fermion in any dimensions
while there are $2^{d}/2$ physical modes in $d$-dimensional naive
overlap fermions. Here the $2^{d}/2$ massless modes in the naive
overlap fermion have the same chiral charge, or equivalently the same
eigenvalue for the ``flavor-chirality'' matrix given by
$\id\otimes(\tau_{3}\otimes\tau_{3})$ in the $d=2$ case. As a
consequence, the flavored Ginsparg-Wilson relations in
(\ref{GW_md_naive}) reduce to unflavored relations for massless modes
as in the staggered case \cite{Adams2}.

This reduction of flavored degrees also affects the index of the Dirac
operators.  We can obtain the indices of the minimally doubled and
naive overlap Dirac operators from the Ginsparg-Wilson relations
(\ref{GW_md_naive}),
\begin{equation}
 \mathrm{Index}(D_{\mathrm{mo}}) = - \frac{1}{2} \mathrm{Tr}
 \left(\frac{H_{\mathrm{md}}}{\sqrt{H_{\mathrm{md}}^2}}\right), \qquad
 \mathrm{Index}(D_{\mathrm{no}}) = - \frac{1}{2} \mathrm{Tr}
 \left(\frac{H_{\mathrm{n}}}{\sqrt{H_{\mathrm{n}}^2}}\right).
\label{OveItheor}
\end{equation}
Now we can easily calculate these quantities analytically: The main
part of the above equations is a sign function as
$H/\sqrt{H^2}=\mathrm{sgn}(H)$.  Thus the trace of this operator
$\mathrm{Tr}\left(\mathrm{sgn}(H)\right)$ gives the difference between
the number of positive and negative eigenvalues at some value of the
mass parameter.  It is essential to fix the mass parameter between
low- and high-lying crossings in the eigenvalue flows of
$H_{\mathrm{md}}$ and $H_{\mathrm{n}}$.  Then the index of the above
overlap Dirac operator becomes just a half of that of the original
Dirac operator,
\begin{equation}
 \mathrm{Index}\left(D_{\mathrm{mo}}\right)
  = \frac{1}{2} \mathrm{Index}\left(D_{\mathrm{md}}\right), \qquad
 \mathrm{Index}\left(D_{\mathrm{no}}\right)
  = \frac{1}{2} \mathrm{Index}\left(D_{\mathrm{n}}\right).
\end{equation}
This relation relies on the property of the Hermitean operator
$\gamma_5 H(m) \gamma_5 = - H(-m)$, as discussed in the end of
Sec.~\ref{sec:SP}. This is also the case with the staggered overlap case.

As shown here, we can construct a single-flavor overlap fermion from
minimally doubled fermions, which however lacks enough discrete
symmetry for a simple continuum limit \cite{MD1, MD2}. On the other
hand, the naive overlap fermion includes lots of flavors (8 for $d=4$)
although it possess hypercubic symmetry. Now we will proceed to obtain
a single-flavor overlap fermion from the naive fermion kernel by
choosing a proper mass term. The key is to utilize the extra flavored
mass terms given in Eqs.~(\ref{exM1})(\ref{exM2}).

For the purpose of constructing a single flavor naive overlap fermion
with the hypercubic symmetry, we make use of the fact that only
species(flavors) with negative mass lead to massless modes in the
overlap formalism.  It means if we introduce a flavored mass term to
assign negative mass to one flavor and positive or zero mass to other
flavors we can derive a single-flavor naive overlap fermion. For
example, in the case of the 2d naive fermion with the original
flavored mass (\ref{2dNMass}), there are two degenerate species with
negative mass. Now we need to lift this degeneracy by adding other
kinds of flavored mass terms. From now we focus on the 2d case for a
while. Now we have two possibilities of flavored mass terms to lift
the degeneracy
\begin{equation}
 M_{\tau_3\otimes\id} = \frac{m_{\tau_3\otimes\id}}{2} \sum_{sym.}
  \left(1+C_1^2\right) C_2,
\end{equation}
\begin{equation}
 M_{\id\otimes\tau_3} = \frac{m_{\id\otimes\tau_3}}{2} \sum_{sym.}
  C_1 \left(1+C_2^2\right),
\end{equation}
which have been already shown in Eqs.~(\ref{exM1})(\ref{exM2}).  Here
we do not consider the mass term of $M_{\id\otimes\id}$ because it
just gives an overall shift of the spectrum.

\begin{figure}
 \begin{center}
  \includegraphics[width=15em]{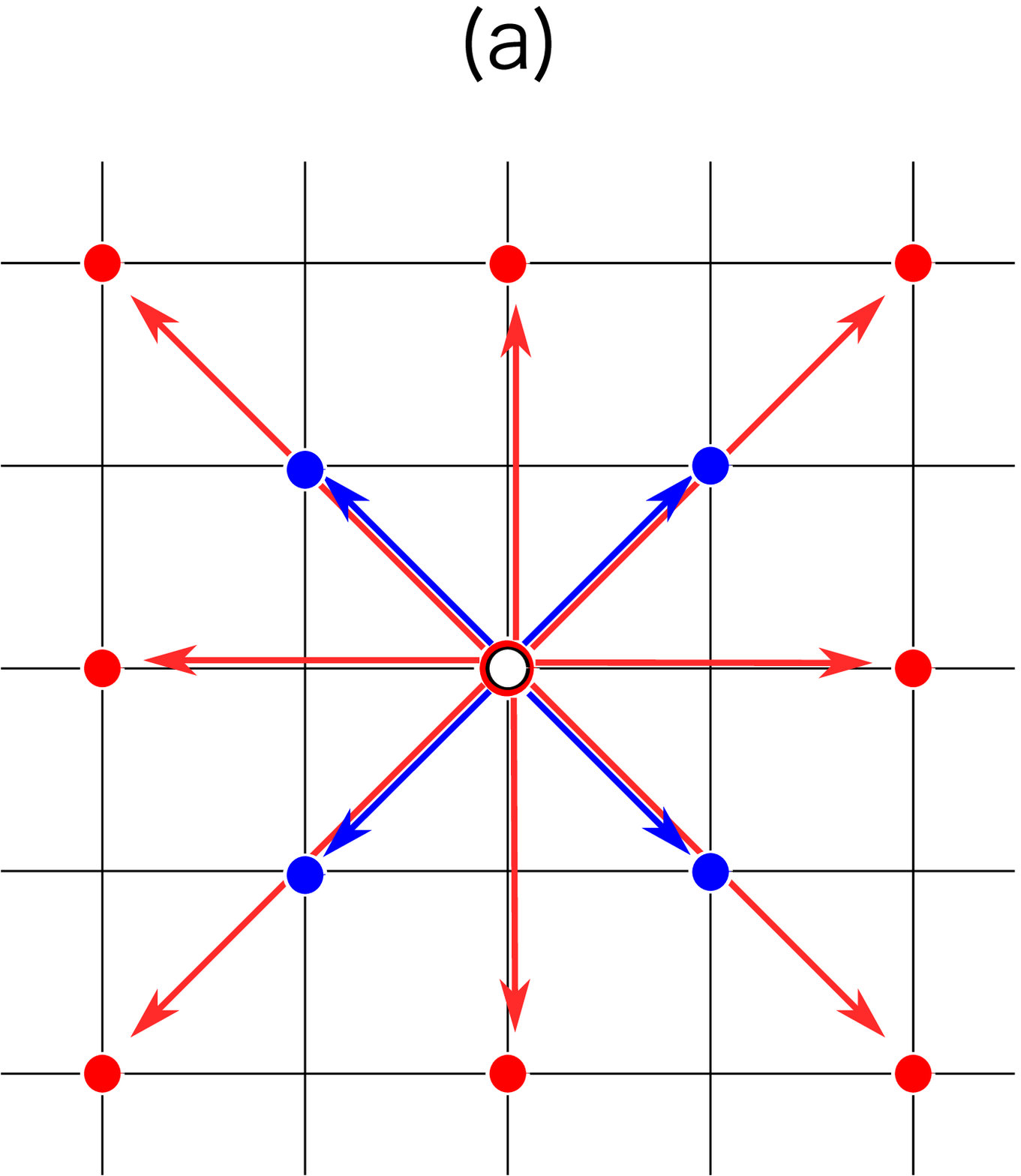} \hspace{5em}
  \includegraphics[width=15em]{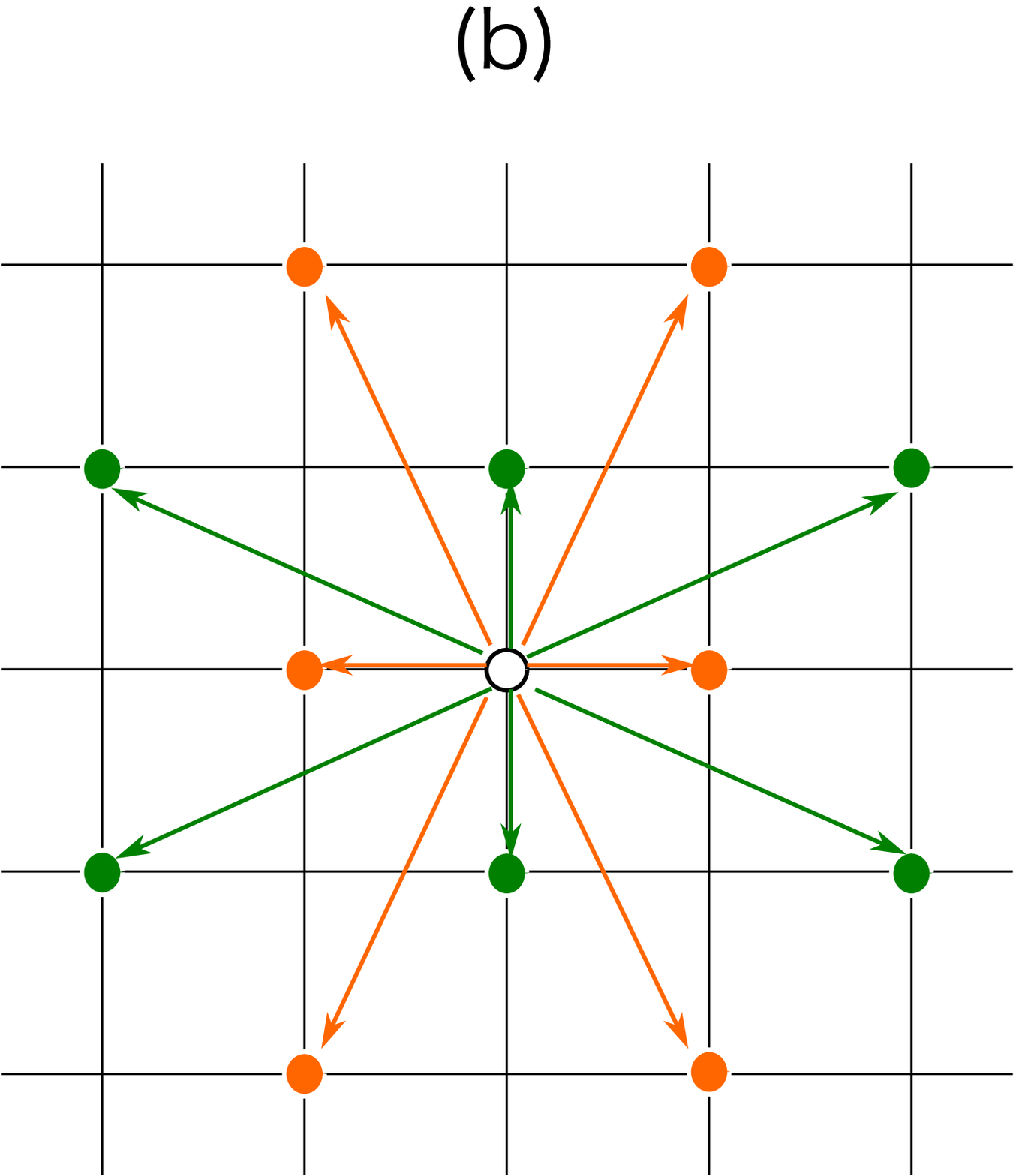}
 \end{center}
 \caption{(Color online) Hopping terms included in naive flavored mass terms:
 (a) isotropic ones, $M_{\tau_3\otimes\tau_3}$ (blue) and
 $M_{\id\otimes\id}$ (red), and (b) anisotropic ones,
 $M_{\tau_3\otimes\id}$ (orange) and $M_{\id\otimes\tau_3}$ (green).}
 \label{NaiveMass_hoppings}
\end{figure}

\begin{figure}
 \begin{center}
  \includegraphics[width=13em]{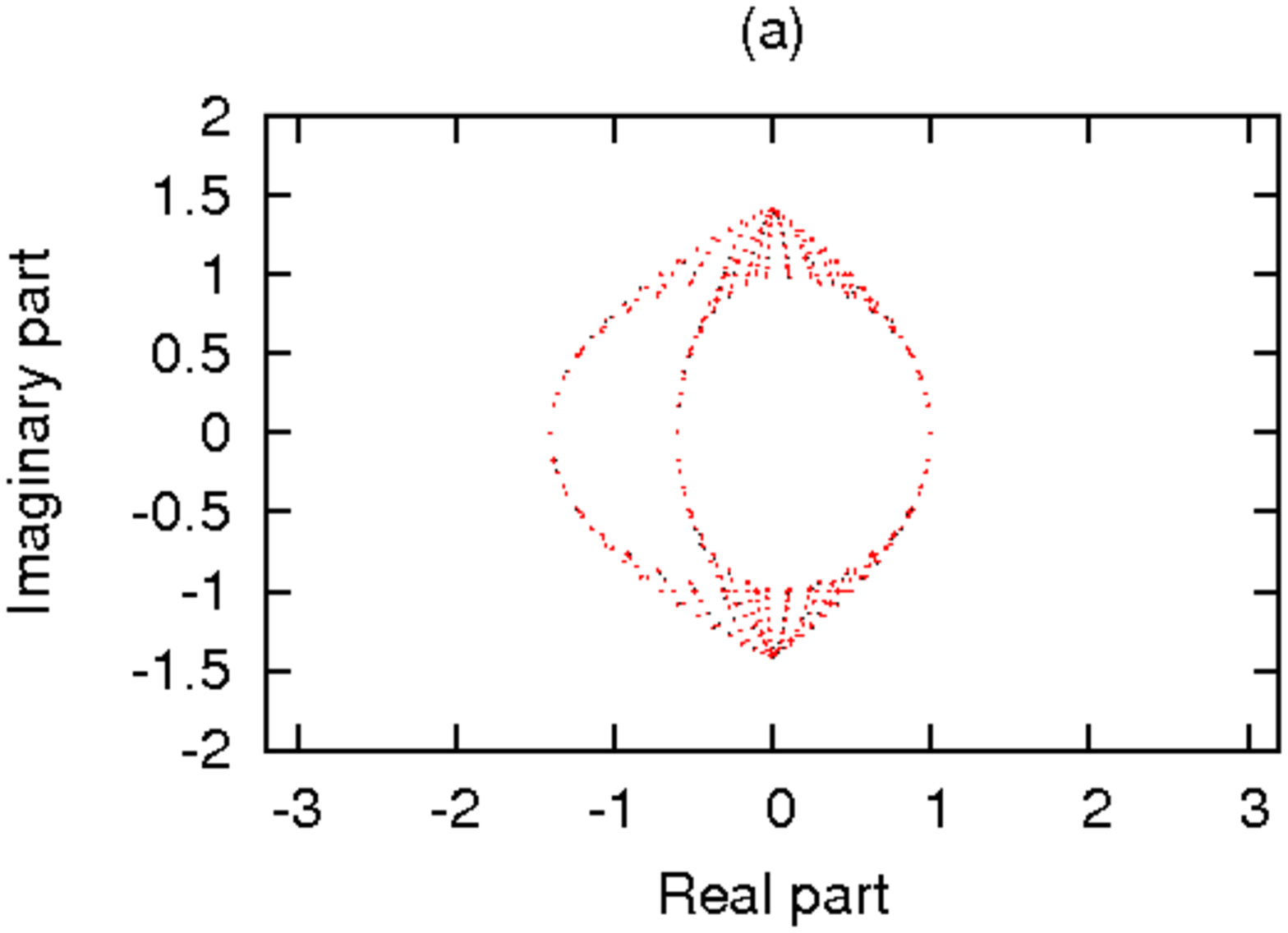}\quad
  \includegraphics[width=13em]{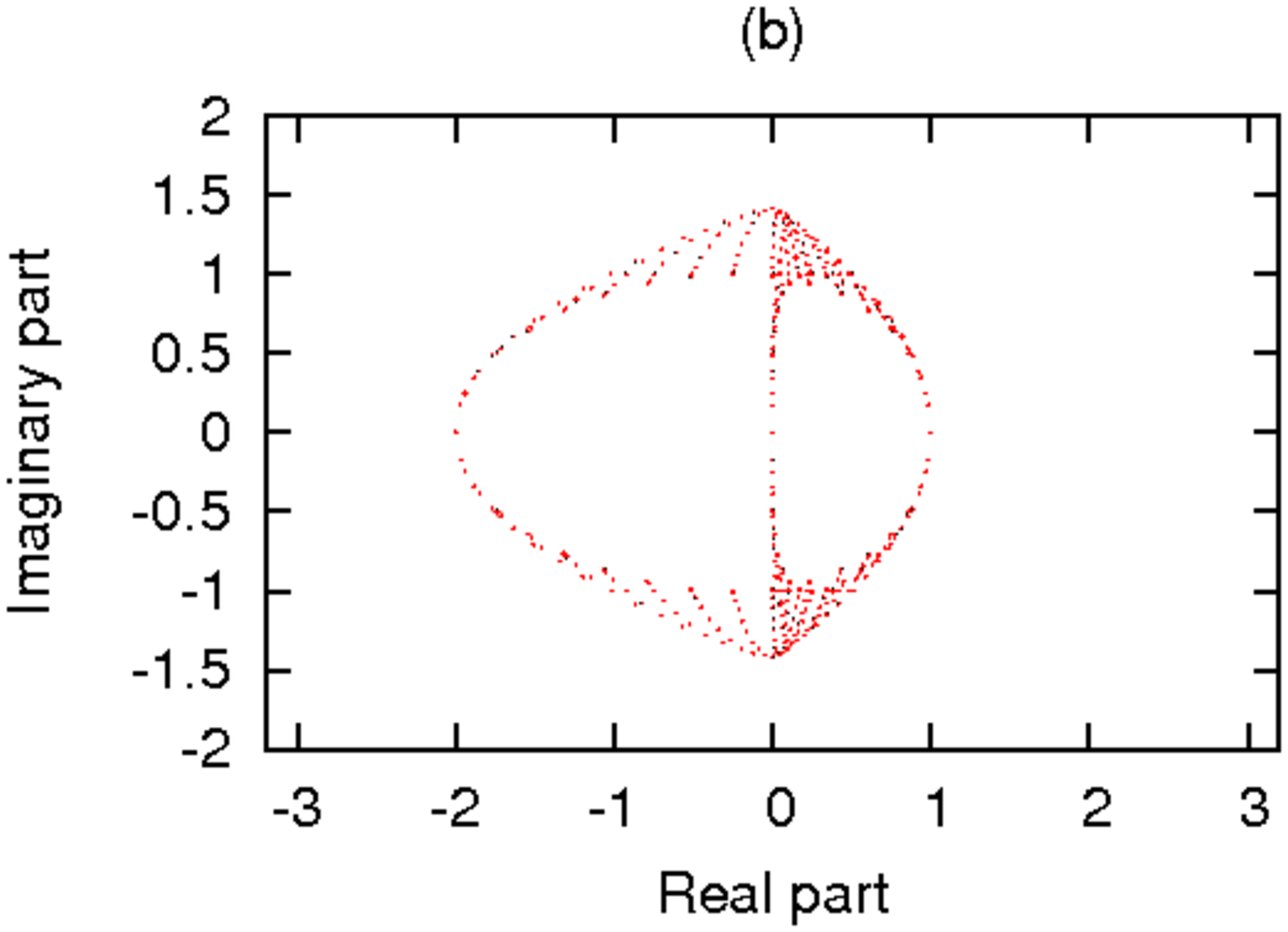}\quad
  \includegraphics[width=13em]{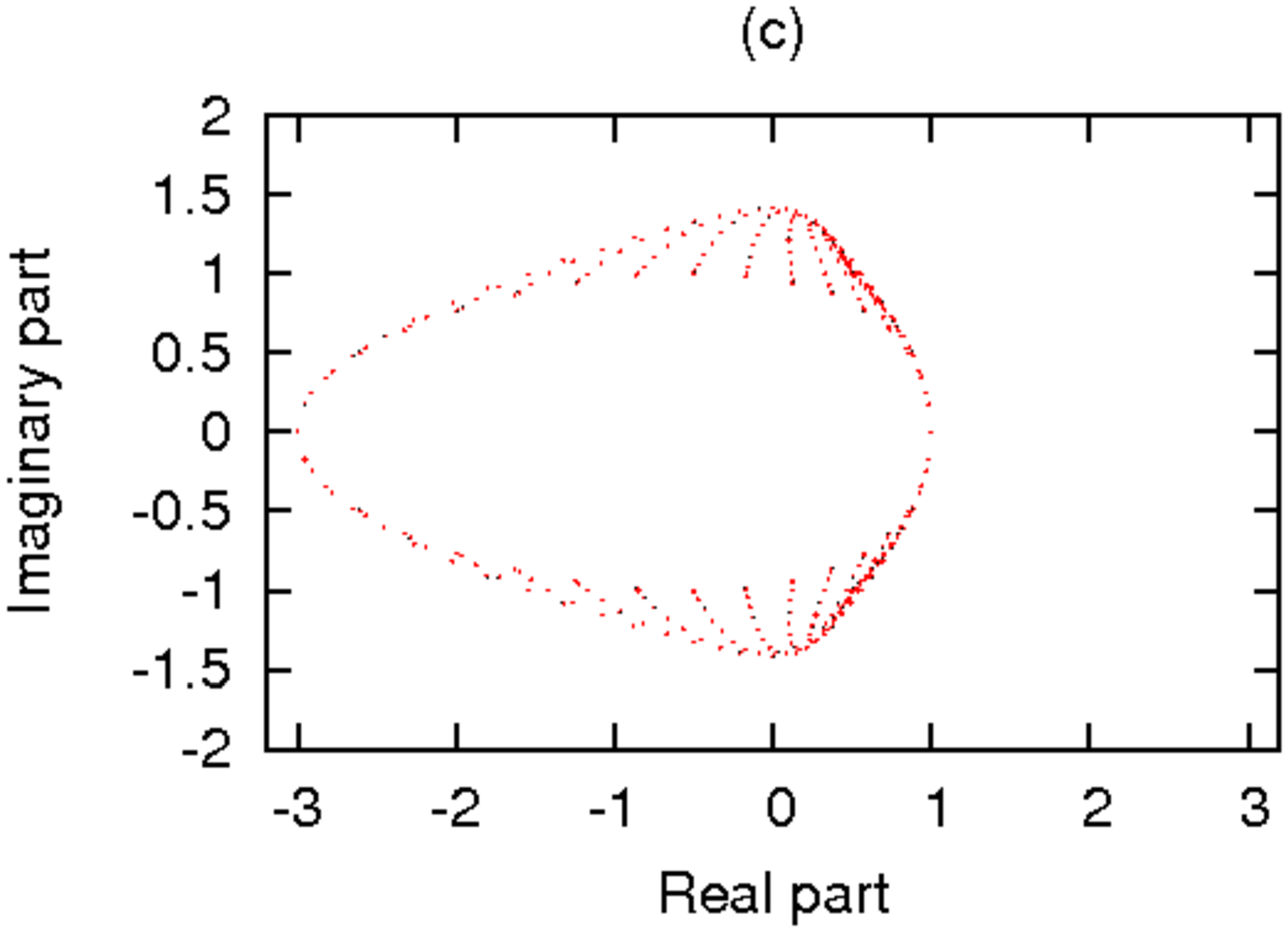}
 \end{center}
 \caption{Complex spectra of the naive Dirac operators for the $d=2$ free
 field case on a $36\times 36$ lattice: (a) $c=0.2$, (b) $c=0.5$ and (c) $c=1$.
 In (a) the doubled negative branch is lifted and one of them goes to the 
 positive direction. In (b) this branch enters a positive range. In (c) 
 it coincides with the positive branch.}
 \label{Dirac_ev_n2d02}
\end{figure}

There is a criterion about how to add these mass terms to the original one 
$M_{\tau_3\otimes\tau_3}$ (\ref{2dNMass}) with preserving the hypercubic symmetry. 
As shown in Fig.~\ref{NaiveMass_hoppings}(b), 
the extra mass terms $M_{\id\otimes\tau_3}$and $M_{\tau_3\otimes\id}$ stand for 
anisotropic hoppings in position space. And the only way to hold hypercubic symmetry is 
add them with the same coefficient to the original one $M_{\tau_3\otimes\tau_3}$,
which has isotropic hoppings as shown in Fig.~\ref{NaiveMass_hoppings}(a).
Thus the possible form of the flavored mass to give single negative mass to the naive 
fermion is given by
\begin{equation}
 -M_\mathrm{n}(c) = -M_{\tau_3\otimes\tau_3} - c
  \left(
   M_{\tau_3\otimes\id} + M_{\id\otimes\tau_3}
  \right).
\end{equation}
where $c$ is an overall coefficient of the two extra mass terms and 
we take account of the convention $D-M$. 
Then we find the simplest case $c=1$ leads to single negative mass as following,
\begin{equation}
 -M_\mathrm{n}(c=1) = \mathrm{diag} (-3, 1, 1, 1)
  \equiv -\tilde M_\mathrm{n}.
\end{equation}
This assigns only one negative mass to the
eigenvalues of the Dirac operator $D_\mathrm{n}-\tilde
M_\mathrm{n}$. Figure \ref{Dirac_ev_n2d02}(c) shows there is a single
negative branch and a tripled positive branch. This means there is only
one flavor with negative mass. We also depict figures for $c=0.2$ 
in Fig.~\ref{Dirac_ev_n2d02}(a) and $c=0.5$ in Fig.~\ref{Dirac_ev_n2d02}(b) 
to convince you that one of the branch is singled out and the
others are tripled.  It describes the situation that the doubled
negative branch is split and one branch goes towards the the positive
branch. Then for the case of $c\sim0.5$ this branch enters a positive
range. Thus it is clear that the negative branch with $c=1$ is singled
out and the positive one should be triply degenerate.

\begin{figure}
 \begin{center}
  \includegraphics[width=18em]{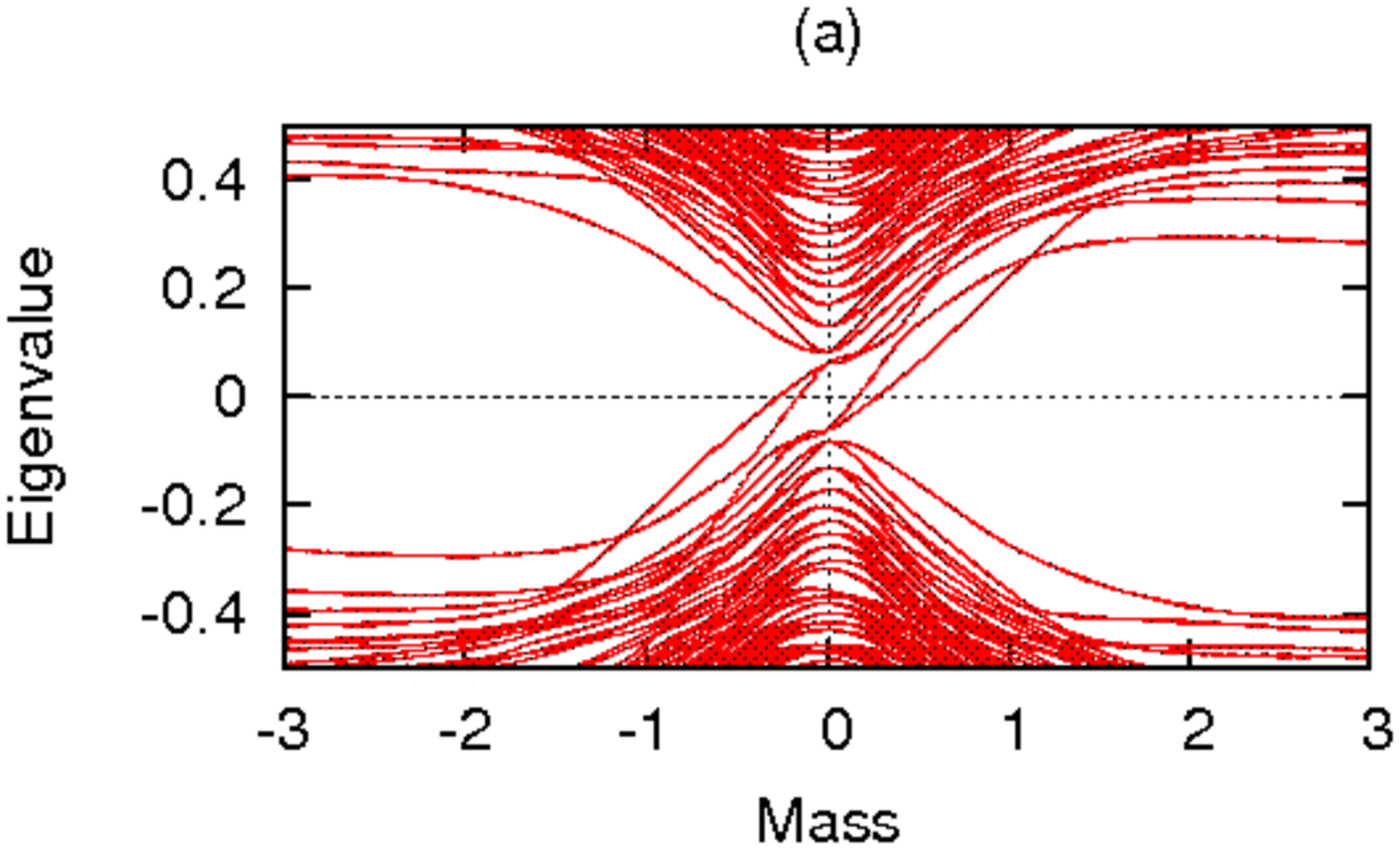} \quad
  \includegraphics[width=18em]{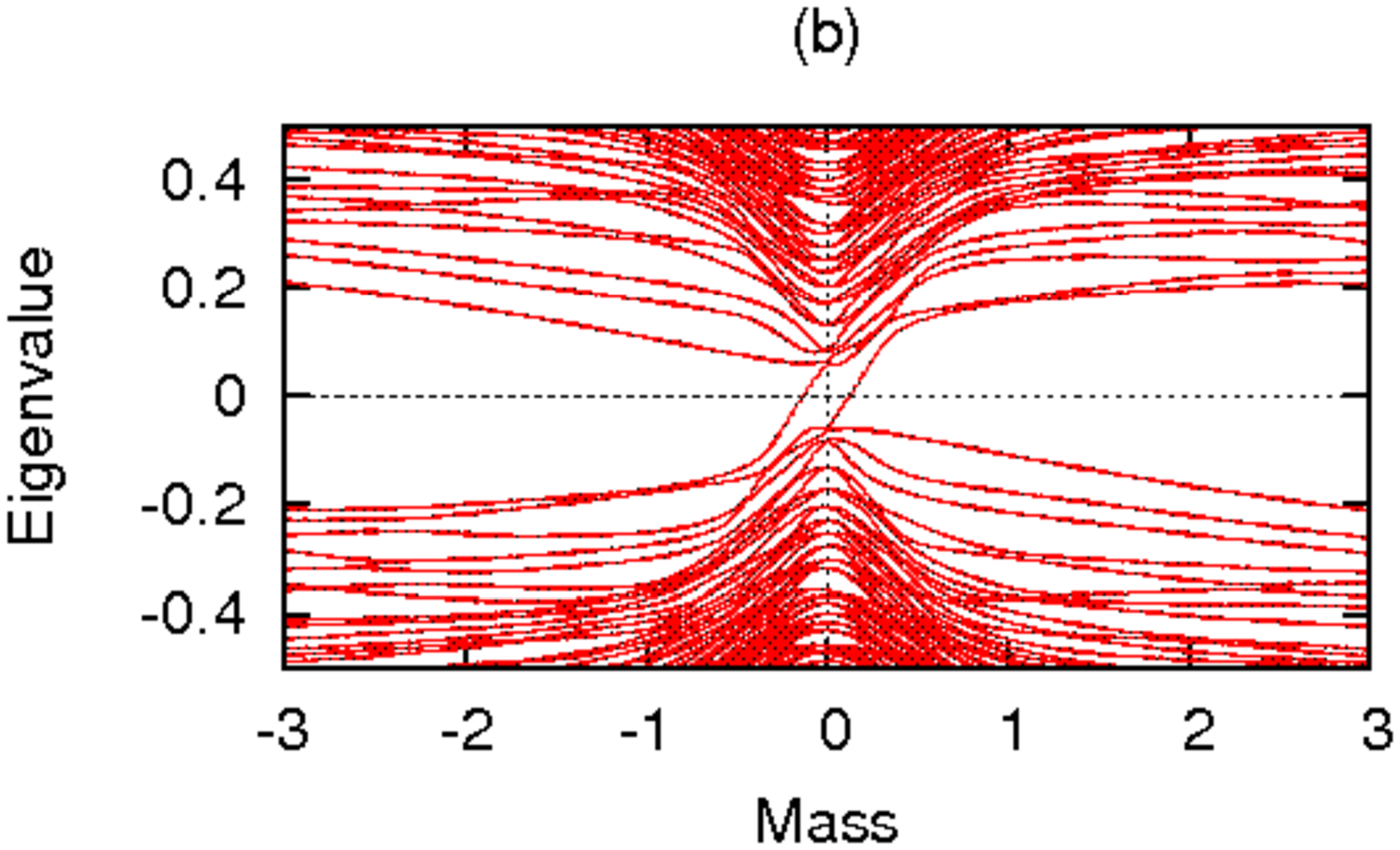} \quad
  \includegraphics[width=18em]{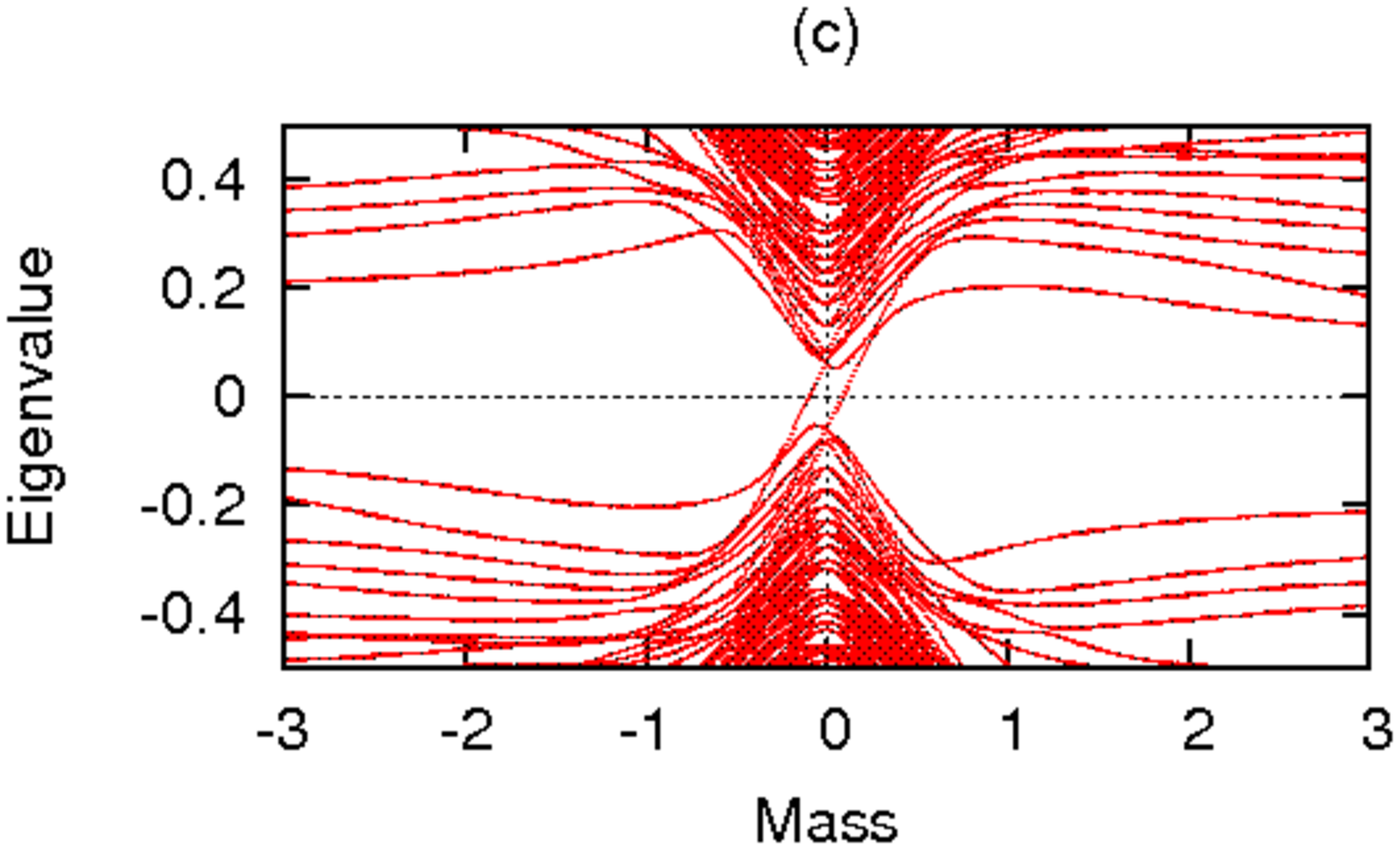}
 \end{center}
 \caption{Spectral flows of the $d=2$ naive Hermitean operators with a
   $Q=1$, $\delta=0.25$ background configuration on a $16\times 16$
   lattice: (a) $c=0.2$, (b) $c=0.5$ and (c) $c=1$.  The doubled flows
   are lifted in (a). Two of the four flows no longer cross zeros in
   (b).  The two single crossings are shown in (c).}
 \label{sf_pic02}
\end{figure}

Now we find one example of the single-flavor naive overlap operator as
following,
\begin{equation}
 \tilde D_{\mathrm{no}} = 1 +
  \frac{D_\mathrm{n} - r \tilde M_\mathrm{n}}
  {\sqrt{(D_\mathrm{n} - r \tilde M_\mathrm{n})^\dag
  (D_\mathrm{n} - r \tilde M_\mathrm{n})}}.
  \label{single_naive_overlap}
\end{equation}
where we introduce an overall mass parameter $r$.  Here we also study
the spectral flow of the Hermitean operator with this mass term.
Indeed it gives us a consistency check between the number of flavors
and the index: Now we have a single flavor overlap fermion
(\ref{single_naive_overlap}), thus the associated index should be
exactly equal to minus the topological charge without a flavor factor
in this dimension. The eigenvalue flow of the naive Hermitean operator
with $c=0.2$, $c=0.5$ and $c=1$ for $Q=1$ are depicted in
Fig.~\ref{sf_pic02}. It is obvious that the doubled flows are separated
in the case of $c=0.2$, and we find two of the four flows no longer
cross zeros for $c=0.5$.  Thus only two crossings remain in the
case of $c=1$. As we have discussed in Eq.~(\ref{OveItheor}), 
the index of the overlap version
is given by minus half difference of positive and negative eigenvalues
of the Hermitean version of the original Dirac operator at some mass
parameter.  It means if you take an overall mass parameter $r$ in
Eq.~(\ref{single_naive_overlap}) between low and high-lying crossings,
the index of exact zero modes of the $d=2$ naive overlap fermion
(\ref{single_naive_overlap}) with $c=1$ is given by $-1$ for $Q=1$.
It generally means
\begin{equation}
 \mathrm{Index}\left(\tilde D_{\mathrm{no}}\right) = - Q,
\end{equation}
which is a quarter of the original index of the 4-species naive
fermion and there is no flavor factor.  We can also check this formula
for other topological charges. Thus we checked the consistency with
the index. In the extension to general dimensions the theorem is given
by
\begin{equation}
 \mathrm{Index}\left(\tilde D_{\mathrm{no}}\right)
  = \frac{1}{2^d} \mathrm{Index}\left(D_{\mathrm{n}}\right)
  = (-1)^{d/2} Q.
  \label{ind_single_naive_overlap}
\end{equation}
Now we can also follow same approach in 4 dimensions, which is slightly more complicated.
Here we do not show details but just present the flavored mass to give only one negative mass
with preserving hypercubic symmetry,
\begin{equation}
 M_{\mathrm{P}} +M_{\mathrm{T}} +M_{\mathrm{V}} +M_{\mathrm{A}},  
\end{equation}
where the meaning of each mass term is shown in Appendix.~\ref{sec:N4d}.
In Fig.~\ref{Dirac_ev_n4d}(c) the eigenvalues of the Dirac operator
with this mass term are depicted.  Here the negative branch is
non-degenerate and the positive one is 15-fold degenerate.  The
associated overlap fermion produces only one massless mode.  Thus the
associated overlap form is a 4d single-flavored naive overlap fermion
with hypercubic symmetry. You can also verify that this 4d version also 
satisfies the theorem in Eq.~(\ref{ind_single_naive_overlap}).

In the end of this section, let us comment on the symmetries and
practical issues of this fermion. As long as the hopping symmetry of
the mass terms is concerned in Fig.~\ref{NaiveMass_hoppings},
hypercubic symmetry is maintained.  Thus, unlike the minimally doubled
case, we do not need to fine-tune additional parameters for the
continuum limit.  Regarding the numerical speed, it should be
computationally comparable to the Wilson overlap. Furthermore the
naive overlap fermion we proposed may have stability of the index against 
disorder due to the original chiral symmetry, although we should also 
take account of large fluctuations and renormalization effects in the
realistic gauge configurations due to non-nearest hopping terms \cite{SmtVnk3}. 
Now we have three examples for overlap varieties: Wilson, Staggered 
and Naive. Which is the best among them is an open question.

%%%%%%%%%%   Summary and Discussion   %%%%%%%%%%

\section{Summary and Discussion}
\label{sec:SD}

In this paper we have shown how the index theorem is realized in naive
and minimally doubled fermions by considering the spectral flow of the
Hermitean version of Dirac operators.  The key is to make use of
a point splitting for flavored mass terms.  We also presented
a new version of overlap fermions composed from the naive fermion
kernel, which is single-flavored and maintains the hypercubic symmetry
essential for a good continuum limit.

In Sec.~\ref{sec:PS} we introduce the point-splitting method to
identify species in momentum space as flavors.  By using this method,
we succeed to define the proper flavored mass terms in minimally
doubled and naive fermions.  In Sec.~\ref{sec:SP} we study the
spectral flow of the Hermitean operators in these fermions. Then it is
shown that the spectral flow correctly illustrates the integer index
determined by gauge field topology both in naive and minimally doubled
fermions up to overall integer factors reflecting the number of
species.  In Sec.~\ref{sec:OF} we study new versions of overlap
fermions composed from these fermion kernels with flavored mass
terms. These fermions satisfy Ginsparg-Wilson relations instead of
usual chiral symmetry.  We show the topological charge can be also
obtained as the index of exact zero modes of the naive and minimally
doubled overlap fermions. Then we show by choosing the flavored mass
term so that only one of species has negative mass, the associated
overlap fermion produces one massless fermionic mode. This
single-flavor naive overlap fermion should possess the hypercubic
symmetry and other properties possessed by the original naive fermion.
Furthermore it may possess more stability for the index against
disorder than Wilson overlap fermion since the naive fermion originally 
has chiral symmetry. However we also need to consider potential large
fluctuations and renormalization effects in the realistic gauge configurations.
Thus further investigation is required to show whether this type of overlap
fermions has advantages for practical simulations over the Wilson overlap fermion.

In the end of this paper, let us discuss realistic possibility to use
this ``Naive Overlap fermion''. Compared to Wilson overlap fermion,
this one begins with only a simple kinetic term, or just a ``naive action''.
Once you fix a flavored mass term such as the one we have shown, to
simulate QCD with this fermion may be easier than with the Wilson overlap.
On the other hand, the staggered overlap fermion is based on
one-component fermionic filed per site while our formulation uses
four-component fields.  This means staggered overlap fermion is likely
to be somewhat better than the naive overlap in terms of numerical
expense. However, to fully answer to this question we need more
detailed numerical research.

%%%%%%%%%%   ACKNOWLEDGMENTS   %%%%%%%%%%

\begin{acknowledgments}
MC is grateful to the Alexander von Humboldt Foundation for support 
for visits to the University of Mainz.  This manuscript has been 
authored under contract number DE-AC02-98CH10886 with the 
U.S.~Department of Energy. Accordingly, the U.S. Government retains
a non-exclusive, royalty-free license to publish or reproduce 
the published form of this contribution, or allow others to do so, 
for U.S.~Government purposes. TK is supported by the JSPS Institutional 
Program for Young Researcher Overseas Visits. TM is supported by 
Grand-in-Aid for the Japan Society for Promotion of Science (JSPS) 
Research Fellows(No.\ 21-1226).

\end{acknowledgments}

%%%%%%%%%%   APPENDIX   %%%%%%%%%%

\appendix
\section{Naive flavored mass in $4$ dimension}\label{sec:N4d}

In this appendix we show details of the point splitting and flavored 
mass terms for the $d=4$ naive fermion.
We introduce 16 point-split fields, corresponding to 16 species
doublers of the $d=4$ naive fermions,
\begin{eqnarray}
 \psi_{(1)}(p-p_{(1)}) & = & \frac{1}{2^4} 
  (1+\cos p_1) (1+\cos p_3) (1+\cos p_3) (1+\cos p_4)
  \Gamma_{(1)} \psi(p), \nonumber \\
 \psi_{(2)}(p-p_{(2)}) & = & \frac{1}{2^4} 
  (1-\cos p_1) (1+\cos p_2) (1+\cos p_3) (1+\cos p_4)
  \Gamma_{(2)} \psi(p), \nonumber \\
 \psi_{(3)}(p-p_{(3)}) & = & \frac{1}{2^4} 
  (1+\cos p_1) (1-\cos p_2) (1+\cos p_3) (1+\cos p_4)
  \Gamma_{(3)} \psi(p), \nonumber \\
 & \vdots & \nonumber \\
 \psi_{(16)}(p-p_{(16)}) & = & \frac{1}{2^4} 
  (1-\cos p_1) (1-\cos p_2) (1-\cos p_3) (1-\cos p_4)
  \Gamma_{(16)} \psi(p),
\end{eqnarray}
where the positions of zeros in the momentum space, chiral charges and definitions 
of transformation matrices $\Gamma_{(i)}$ are listed in Table \ref{gammas_naive4d}.
Here a set of gamma matrices $\gamma_{\mu}^{(i)}$ defined for each zero is given 
by this $\Gamma_{(i)}$ as $\Gamma^{-1}_{(i)}\gamma_{\mu}\Gamma_{(i)}=\gamma_{\mu}^{(i)}$. 
We classify these zeros depending on this $\Gamma_{(i)}$.
For example we denote A: axial for the case of $\Gamma_{(2)}=i\gamma_{1}\gamma_{5}$
while we assign T: Tensor for $\Gamma_{(4)}=i\gamma_{1}\gamma_{2}$.
\begin{table}[tbp]
 \begin{center}
  {\tabcolsep = 1em
  \begin{tabular}{ccccc} \hline\hline
   label & position & $\chi$ charge & $\Gamma$ & type \\ \hline
   $1$ & $(0, 0, 0, 0)$ & $+$ & $\id$ & S \\
   $2$ & $(\pi, 0, 0, 0)$ & $-$ & $i\gamma_1\gamma_5$ & A \\
   $3$ & $(0, \pi, 0, 0)$ & $-$ & $i\gamma_2\gamma_5$ & A \\
   $4$ & $(\pi, \pi, 0, 0)$ & $+$ & $i\gamma_1\gamma_2$ & T \\
   $5$ & $(0, 0, \pi, 0)$ & $-$ & $i\gamma_3\gamma_5$ & A \\
   $6$ & $(\pi, 0, \pi, 0)$ & $+$ & $i\gamma_1\gamma_3$ & T \\
   $7$ & $(0, \pi, \pi, 0)$ & $+$ & $i\gamma_2\gamma_3$ & T \\
   $8$ & $(\pi, \pi, \pi, 0)$ & $-$ & $\gamma_4$ & V \\ 
   $9$ & $(0, 0, 0, \pi)$ & $-$ & $i\gamma_4\gamma_5$ & A \\
   $10$ & $(\pi, 0, 0, \pi)$ & $+$ & $i\gamma_1\gamma_4$ & T \\
   $11$ & $(0, \pi, 0, \pi)$ & $+$ & $i\gamma_2\gamma_4$ & T \\
   $12$ & $(\pi, \pi, 0, \pi)$ & $-$ & $\gamma_3$ & V \\
   $13$ & $(0, 0, \pi, \pi)$ & $+$ & $i\gamma_3\gamma_4$ & T \\
   $14$ & $(\pi, 0, \pi, \pi)$ & $-$ & $\gamma_2$ & V \\
   $15$ & $(0, \pi, \pi, \pi)$ & $-$ & $\gamma_1$ & V \\
   $16$ & $(\pi, \pi, \pi, \pi)$ & $+$ & $\gamma_5$ & P \\ \hline\hline
  \end{tabular}
  }
 \end{center}
 \caption{Positions of zeros, chiral charges and definitions of transformation matrices
 for the $d=4$ naive fermions. Letters of S, V, T, V and P stand for
 Scalar, Vector, Tensor, Axial-vector and Pseudo-scalar, respectively.}
 \label{gammas_naive4d}
\end{table}
We introduce a flavor multiplet field with 16 components as
\begin{equation}
 \Psi(p) = 
  \left(
   \begin{array}{c}
       \psi_{(1)}(p-p_{(1)}) \\ \psi_{(2)}(p-p_{(2)})
	\\ \vdots \\ \psi_{(16)}(p-p_{(16)})
   \end{array}
  \right).
\end{equation}
Then 16 flavored mass terms are given by
\[
{\renewcommand{\arraystretch}{2}
 \begin{array}{ccccl}
  \mathrm{S} & : & \bar\Psi
   \left(
    \id\otimes\left(\id\otimes\id\otimes\id\otimes\id\right)
   \right)\Psi
   &=& \displaystyle \frac{1}{2^4} 
   \left(
    \prod_{\mu=1}^4 \left(1+\cos^2p_\mu\right)
   \right) \bar\psi\psi \\
  \mathrm{V} & : & \bar\Psi
   \left(
    \id\otimes\left(\tau_3\otimes\id\otimes\id\otimes\id\right)
   \right)\Psi
   &=& \displaystyle \frac{1}{2^3} \cos p_1 
   \left(
    \prod_{\mu=2}^4 \left(1+\cos^2p_\mu\right)
    \right) \bar\psi\psi \\
  \mathrm{T} & : & \bar\Psi
   \left(
    \id\otimes\left(\tau_3\otimes\tau_3\otimes\id\otimes\id\right)
   \right)\Psi
   &=& \displaystyle \frac{1}{2^2} \cos p_1 \cos p_2 
   \left(
    \prod_{\mu=3}^4 \left(1+\cos^2 p_\mu\right)
   \right) \bar\psi\psi\\
  \mathrm{A} & : & \bar\Psi
   \left(
    \id\otimes\left(\id\otimes\tau_3\otimes\tau_3\otimes\tau_3\right)
   \right)\Psi
   &=& \displaystyle \frac{1}{2} (1+\cos^2p_1) 
   \left(
    \prod_{\mu=2}^4 \cos p_\mu
   \right) \bar\psi\psi\\
  \mathrm{P} & : & \bar\Psi
   \left(
    \id\otimes\left(\tau_3\otimes\tau_3\otimes\tau_3\otimes\tau_3\right)
   \right)\Psi
   &=& \displaystyle 
   \left(
    \prod_{\mu=1}^4 \cos p_\mu
   \right) \bar\psi\psi\\
 \end{array}}
 \qquad \mbox{etc},
\]
where we introduce a multiplet representation as four direct products
of the Pauli matrix to express the 16 flavor structure. In this
representation the chiral transformation matrix $\gamma_{5}$ is
converted to
$\gamma_{5}\otimes(\tau_3\otimes\tau_3\otimes\tau_3\otimes\tau_3)$,
under which the naive action is invariant.  Here the V-type includes 4
varieties while the T-type has 6 varieties and the A-type 4 varieties,
giving then the total number as 16. Each of these varieties breaks the
hypercubic symmetry of the lattice since their hoppings are
anisotropic. Thus we need to take a proper combination of them in
order to construct a flavored mass term with hypercubic symmetry. For
example we consider the sum of the 4 varieties with the same ratios
for the V-type as in the case of $d=2$.  It is also the case with
the T- and A-types.  Thus the 5 types of flavored masses with the
symmetry in terms of the original fermion field are given by
\begin{equation}
 M_{\mathrm{S}} = \frac{m_{\mathrm{S}}}{2^4} \sum_{sym.} 
  \prod_{\mu=1}^4 \left( 1+C_\mu^2 \right),
\end{equation}
\begin{equation}
 M_{\mathrm{V}} = \frac{m_{\mathrm{V}}}{2^3} \sum_{perm.} \sum_{sym.}
  C_\mu \prod_{\nu(\not=\mu)} \left( 1+C_\nu^2 \right), 
\end{equation}
\begin{equation}
 M_{\mathrm{T}} = \frac{m_{\mathrm{T}}}{2^2} \sum_{perm.} \sum_{sym.}
  C_\mu C_\nu \left( 1+C_\rho^2 \right) \left( 1+C_\sigma^2\right), 
\end{equation}
\begin{equation}
 M_{\mathrm{A}} = \frac{m_{\mathrm{A}}}{2} \sum_{perm.} \sum_{sym.}
  \left(1+C_\mu^2\right) \prod_{\nu(\not=\mu)} C_\nu,
\end{equation}
\begin{equation}
 M_{\mathrm{P}} = m_{\mathrm{P}} \sum_{sym.} 
  \prod_{\mu=1}^4 C_\mu,
\label{P-mass}  
\end{equation}
where $\sum_{perm.}$ means summation over permutations of the
space-time indices.

\begin{figure}
 \begin{center}
  \includegraphics[height=13em]{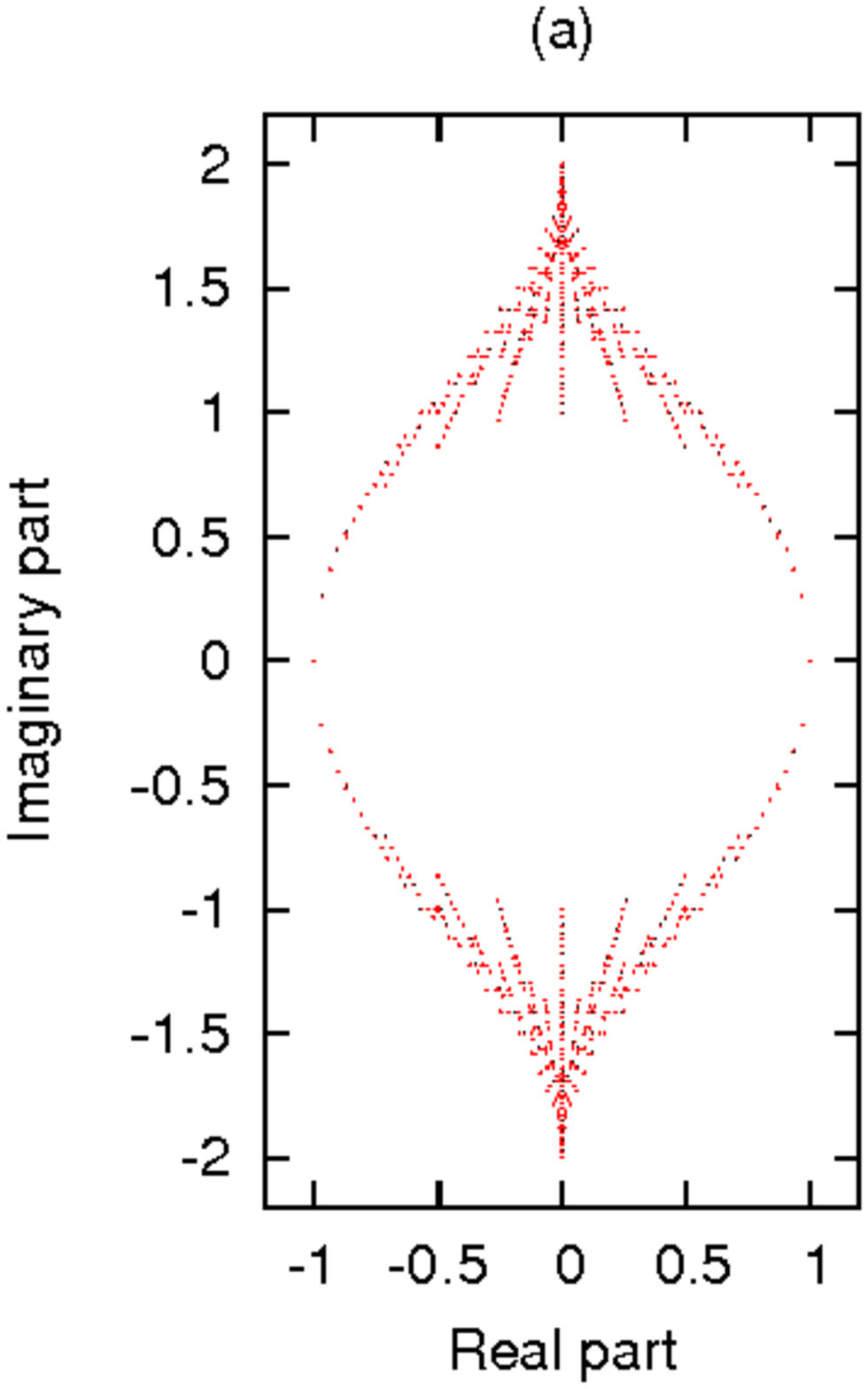}\quad
  \includegraphics[height=13em]{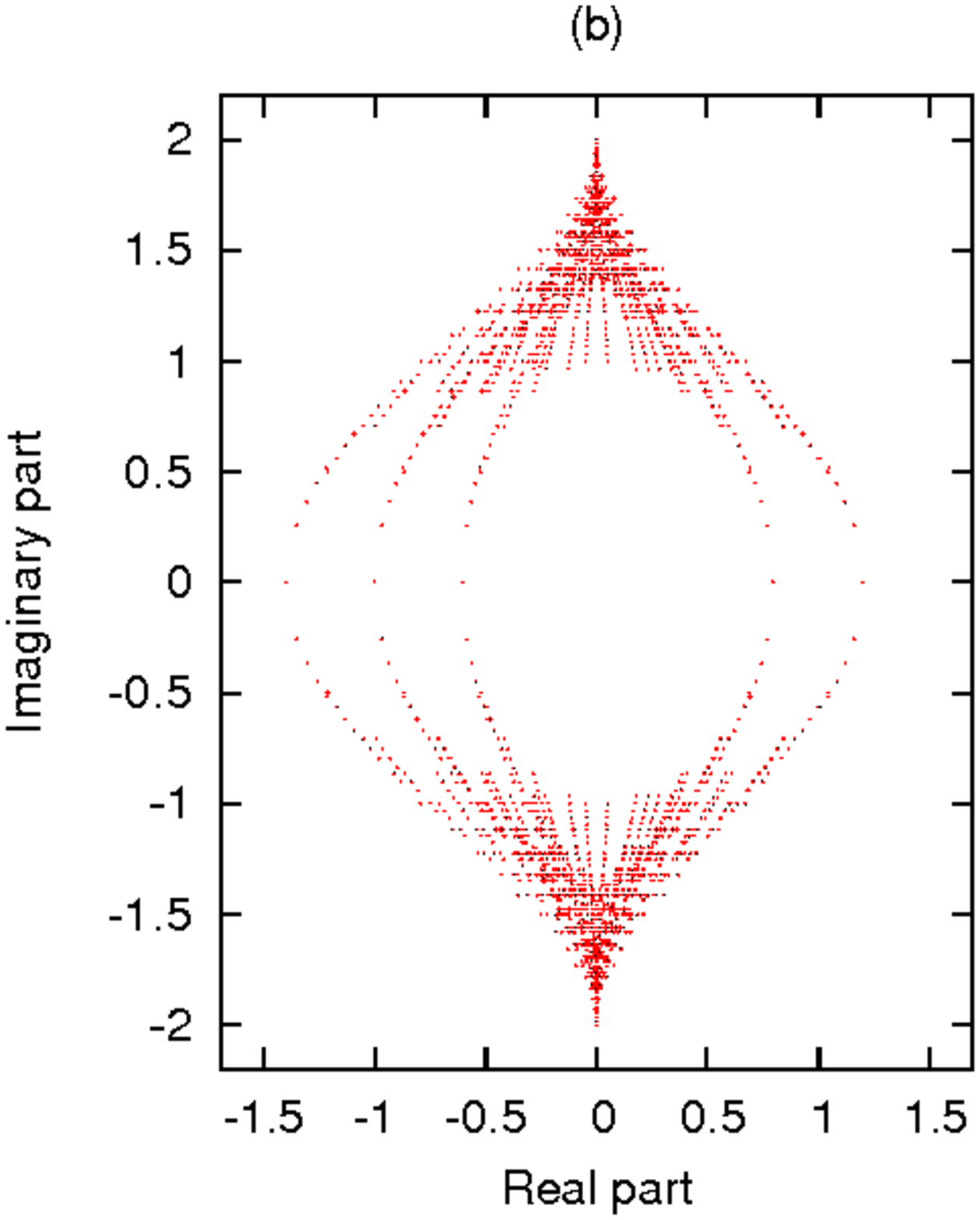}\quad
  \includegraphics[height=13em]{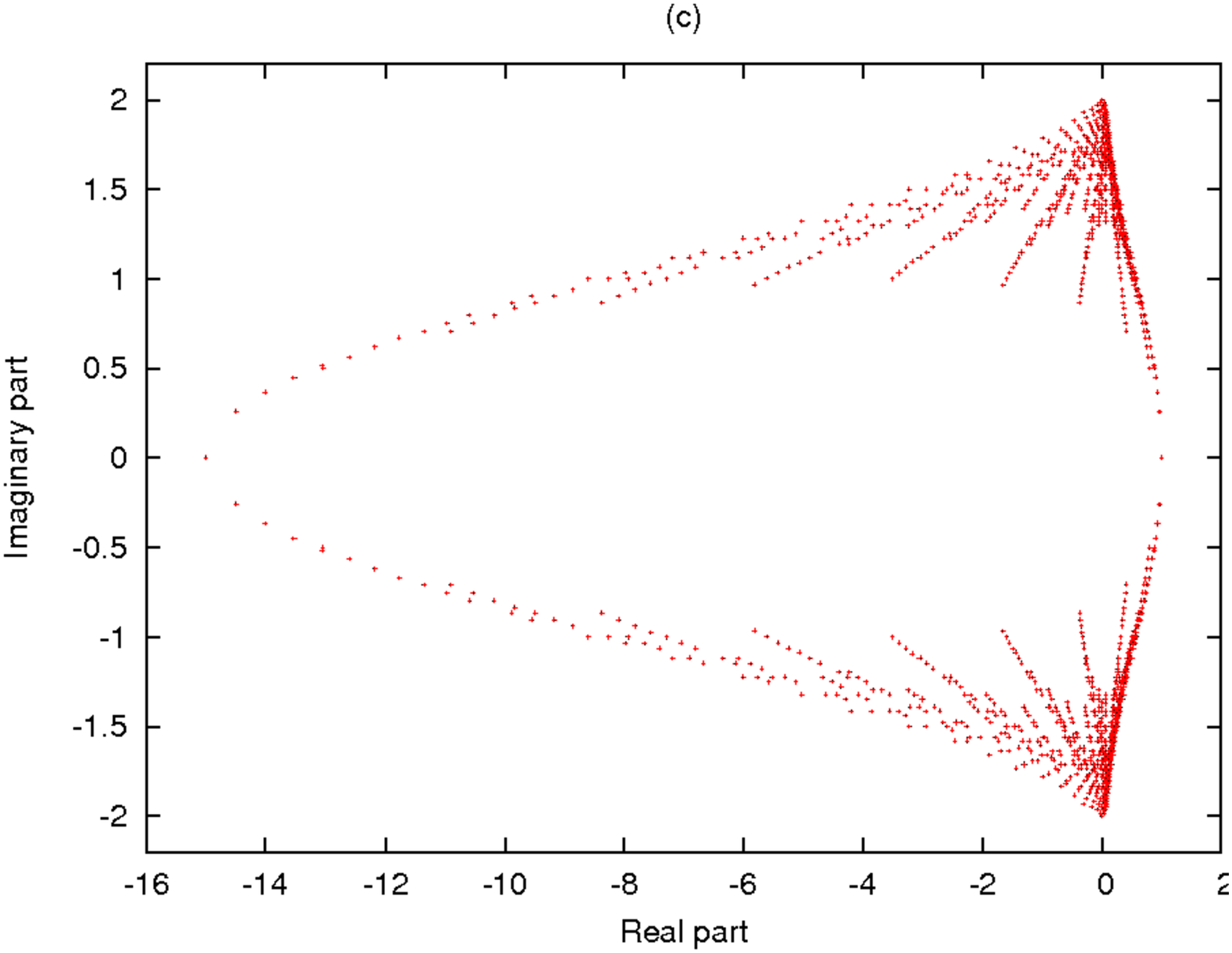}
 \end{center}
 \caption{Complex spectra of non-Hermitean Dirac operators
 for the $d=4$ free field case in momentum space
 with $16^{4}$ grids of the brillouin zone.
 (a) $D_{\mathrm{n}}- M_{\mathrm{P}}$. 
 (b) $D_{\mathrm{n}}-(M_{\mathrm{P}}+0.1M_{\mathrm{A}})$.
 (c) $D_{\mathrm{n}}-(M_{\mathrm{P}}+M_{\mathrm{V}} +M_{\mathrm{T}} +M_{\mathrm{A}})$.}
 \label{Dirac_ev_n4d}
\end{figure}

Now we derive the flavored mass terms required to detect the index from
the spectral flow of the Hermitean operator. As in the $d=2$ case, it
should be constructed so that the associated Hermitean operator has
a flavor-singlet mass part as 
$\gamma_{5}M \sim \gamma_{5}\otimes(\id\otimes\id\otimes\id\otimes\id)$.
Such a mass term is just the P-type mass (\ref{P-mass}).
Thus the flavored mass term for the Hermitean operator is given by
\begin{equation}
 M_{\mathrm{P}} = m_{\mathrm{P}} \sum_{sym.} 
  \prod_{\mu=1}^4 C_\mu.  
  \label{4dIndmass}
\end{equation}
With the Hermitean operator $H_{\rm n}=\gamma_{5}(D_{\mathrm n}-M_{\mathrm{P}})$,
we reveal the index theorem with the naive fermion as in the $d=2$ case.
Here we only show the figure for eigenvalues of the free Dirac operator
$D_{\rm n}-M_{\rm P}$ in Fig.~\ref{Dirac_ev_n4d}(a). The mass term splits the modes into two branches, 
which are 8 fold degenerate. If we introduce other types of mass terms, 
the degeneracy is lifted as seen in Fig.~\ref{Dirac_ev_n4d}(b).

Next we show the flavored mass term to yield a single-flavor naive overlap
fermion in $4$d. As in the case of $2$d there are some possibilities to
realize it. The simplest example of the mass term to yield a single-flavor 
naive overlap fermion with hypercubic symmetry is given by
\begin{equation}
 M_{\mathrm{P}} +M_{\mathrm{V}} +M_{\mathrm{T}} +M_{\mathrm{A}}.  
\label{4dSingmass}
\end{equation}
The eigenvalues of the Dirac operator with this mass term is depicted
in Fig.~\ref{Dirac_ev_n4d}(c). Here the negative branch is single and
the positive one is 15 fold degenerate. It means the associated
overlap fermion produces only one massless mode. Furthermore, as seen
from the construction, it should possess hypercubic symmetry. Thus it
could be a useful fermion action for lattice QCD simulation.

We now comment on some connections with other works.
The P-type mass term actually corresponds to the modified gamma
matrix $\Gamma_5 = (\prod_{\mu=1}^4 \cos p_\mu) \gamma_5$, which is
introduced in Ref.~\cite{CreutzN} to study the effects of topological 
charge with the $4$d naive fermion. In some sense we have given its 
theoretical foundation by the point-splitting method.
It is also similar to $\Gamma_5\sim\gamma_5\otimes\id$ of
the staggered fermions \cite{SmtGlt}.
Regarding the T-type masses, they are similar to the flavored mass
terms for the single-flavor staggered overlap fermion proposed 
in \cite{Hoel}. These similarities are reasonable since the staggered
fermions can be obtained from the naive fermions \cite{Sha}.

%%%%%%%%%%   References   %%%%%%%%%%

\end{document}